\documentclass[preprint2]{aastex62}
\usepackage{color}
\usepackage{natbib}
\usepackage{hyperref}
\graphicspath{{./}{Figures/}}
\newcommand{\HII}{\textrm{H~{\textsc{ii}}}}
\newcommand{\HI}{\textrm{H~{\textsc{i}}}}
\newcommand{\CII}{\textrm{C~{\textsc{ii}}}}
\newcommand{\OI}{\textrm{O~{\textsc{i}}}}
\newcommand{\FeII}{\textrm{Fe~{\textsc{ii}}}}
\newcommand{\FeI}{\textrm{Fe~{\textsc{i}}}}
\newcommand{\ArIII}{\textrm{Ar~{\textsc{iii}}}}
\newcommand{\ArII}{\textrm{Ar~{\textsc{ii}}}}
\newcommand{\NeIII}{\textrm{Ne~{\textsc{iii}}}}
\newcommand{\NeII}{\textrm{Ne~{\textsc{ii}}}}
\newcommand{\SIV}{\textrm{S~{\textsc{iv}}}}
\newcommand{\SIII}{\textrm{S~{\textsc{iii}}}}
\newcommand{\SI}{\textrm{S~{\textsc{i}}}}
\newcommand{\SII}{\textrm{S~{\textsc{ii}}}}
\newcommand{\PIII}{\textrm{P~{\textsc{iii}}}}
\newcommand{\NiII}{\textrm{Ni~{\textsc{ii}}}}
\newcommand{\FI}{\textrm{F~{\textsc{i}}}}
\newcommand{\ClI}{\textrm{Cl~{\textsc{i}}}}

\newcommand{\red}[1]{\textcolor{red}{#1}}

\shorttitle{PDRs4All}
\shortauthors{Bern\'{e}, Habart, Peeters \& PDRs4All team}

\begin{document}

\title{PDRs4All: A JWST Early Release Science Program on radiative feedback from massive stars}

\correspondingauthor{PI Team: Bern\'{e}, Habart, \& Peeters}
\email{pis@jwst-ism.org}

\author{PI Team:}
\affiliation{}

\author[0000-0002-1686-8395]{Olivier Bern\'{e}}
\affiliation{Institut de Recherche en Astrophysique et Plan\'etologie, Universit\'e de Toulouse, CNRS, CNES, UPS, 9 Av. du colonel Roche, 31028 Toulouse Cedex 04, France}

\author[0000-0001-9136-8043]{\'{E}milie Habart}
\affil{Institut d'Astrophysique Spatiale, Universit\'e Paris-Saclay, CNRS,  B$\hat{a}$timent 121, 91405 Orsay Cedex, France}

\author[0000-0002-2541-1602]{Els Peeters}
\affiliation{Department of Physics \& Astronomy, The University of Western Ontario, London ON N6A 3K7, Canada}
\affiliation{Institute for Earth and Space Exploration, The University of Western Ontario, London ON N6A 3K7, Canada}
\affiliation{Carl Sagan Center, SETI Institute, 339 Bernardo Avenue, Suite 200, Mountain View, CA 94043, USA}

\author{Core Team:}
\affiliation{}

\author[0000-0003-2364-2260]{Alain Abergel}
\affil{Institut d'Astrophysique Spatiale, Universit\'e Paris-Saclay, CNRS,  B$\hat{a}$timent 121, 91405 Orsay Cedex, France}

\author[0000-0003-4179-6394]{Edwin A. Bergin}
\affil{Department of Astronomy, University of Michigan, 1085 South University Avenue, Ann Arbor, MI 48109, USA}

\author[0000-0002-8452-8675]{Jeronimo Bernard-Salas}
\affil{ACRI-ST, Centre d'Etudes et de Recherche de Grasse (CERGA), 10 Av. Nicolas Copernic, 06130 Grasse, France}

\author[0000-0003-1532-7818]{Emeric Bron}
\affil{LERMA, Observatoire de Paris, PSL Research University, CNRS, Sorbonne Universit\'es, F-92190 Meudon, France}

\author[0000-0002-2666-9234]{Jan Cami}
\affiliation{Department of Physics \& Astronomy, The University of Western Ontario, London ON N6A 3K7, Canada}
\affiliation{Institute for Earth and Space Exploration, The University of Western Ontario, London ON N6A 3K7, Canada}
\affiliation{Carl Sagan Center, SETI Institute, 339 Bernardo Avenue, Suite 200, Mountain View, CA 94043, USA}

% ...


\begin{abstract}

Massive stars disrupt their natal molecular cloud material through radiative and mechanical feedback processes. These processes have profound effects on the evolution of interstellar matter in our Galaxy and throughout the Universe, from the era of vigorous star formation at redshifts of 1-3 to the present day. 
The dominant feedback processes can be probed by observations of the Photo-Dissociation Regions (PDRs) where the far-ultraviolet photons of massive stars create warm regions of gas and dust in the neutral atomic and molecular gas.
PDR emission provides a unique tool to study in detail the physical and chemical processes that are relevant for most of the mass in inter- and circumstellar media including diffuse clouds, proto-planetary disks and molecular cloud surfaces, globules, planetary nebulae, and star-forming  regions. PDR emission dominates the infrared (IR) spectra of star-forming galaxies. Most of the Galactic and extragalactic observations obtained with the James Webb Space Telescope (JWST) will therefore arise in PDR emission. In this paper we present an Early Release Science program using the MIRI, NIRSpec, and NIRCam instruments dedicated to the observations of an emblematic and nearby PDR: the Orion Bar. These early JWST observations will provide template datasets designed to identify key PDR characteristics in JWST observations. These data will serve to benchmark PDR models and extend them into the JWST era. We also present the Science-Enabling products that we will provide to the community. These template datasets and Science-Enabling products will guide the preparation of future proposals on star-forming regions in our Galaxy and beyond and will facilitate data analysis and interpretation of forthcoming JWST observations. \\
\end{abstract}

\section{Introduction}
The James Webb Space Telescope (JWST, \citealt{gardner2006}) is a 6.5m space telescope launched in December 2021 and is developed by the National Aeronautics and Space Administration (NASA), the European Space Agency (ESA), and the Canadian Space Agency (CSA). Following recommendations of the science advisory board of the JWST,  the Space Telescope Science Institute (STScI), in charge of the scientific operations of the JWST, issued a call for ``\href{https://jwst.stsci.edu/science-planning/calls-for-proposals-and-policy/early-release-science-program}{Early Release Science}'' (ERS) programs. 
The goals of these programs are 1) to provide first-look public data to the astronomical community as soon as possible after launch, 2) to further test the instruments and observing modes in addition to tests performed during commissioning and showcase the technical capabilities of JWST, and 3) to help prepare the community for General Observers (GO) proposals. An important aspect of ERS programs is that they must deliver highly-processed data quickly (within 3 to 5 months of observations) and provide Science-Enabling Products (SEPs) to the community. 

 In this paper, we present one of the 13 accepted ERS programs called ``PDRs4All: Radiative feedback from massive stars" (ID1288)\footnote{\url{pdrs4all.org}; \url{https://www.stsci.edu/jwst/science-execution/program-information.html?id=1288}} which is dedicated to studying the interactions of massive stars with their surroundings. We first describe the general scientific context in Section \ref{sec:context}. We then describe the immediate objectives of this program in Section~\ref{sec:immediate-goals} and its science objectives in Section~\ref{sec:science-objectives}.
 We discuss the target, the Orion Bar, in Section~\ref{sec:target} and describe simulated infrared (IR) spectra for this source in Section~\ref{sec:simulated-spectra}. The planned observations are presented in Section~\ref{sec:obs}.  Section~\ref{sec:SEPs} gives an overview of the Science-Enabling Products developed for this program. We briefly present the team in Section~\ref{sec:team} and conclude in Section~\ref{sec:conclusions}.

\section{Importance of PDRs in the JWST era }
\label{sec:context}

\begin{figure*}[!t]
    \centering
    \includegraphics[scale=0.60]{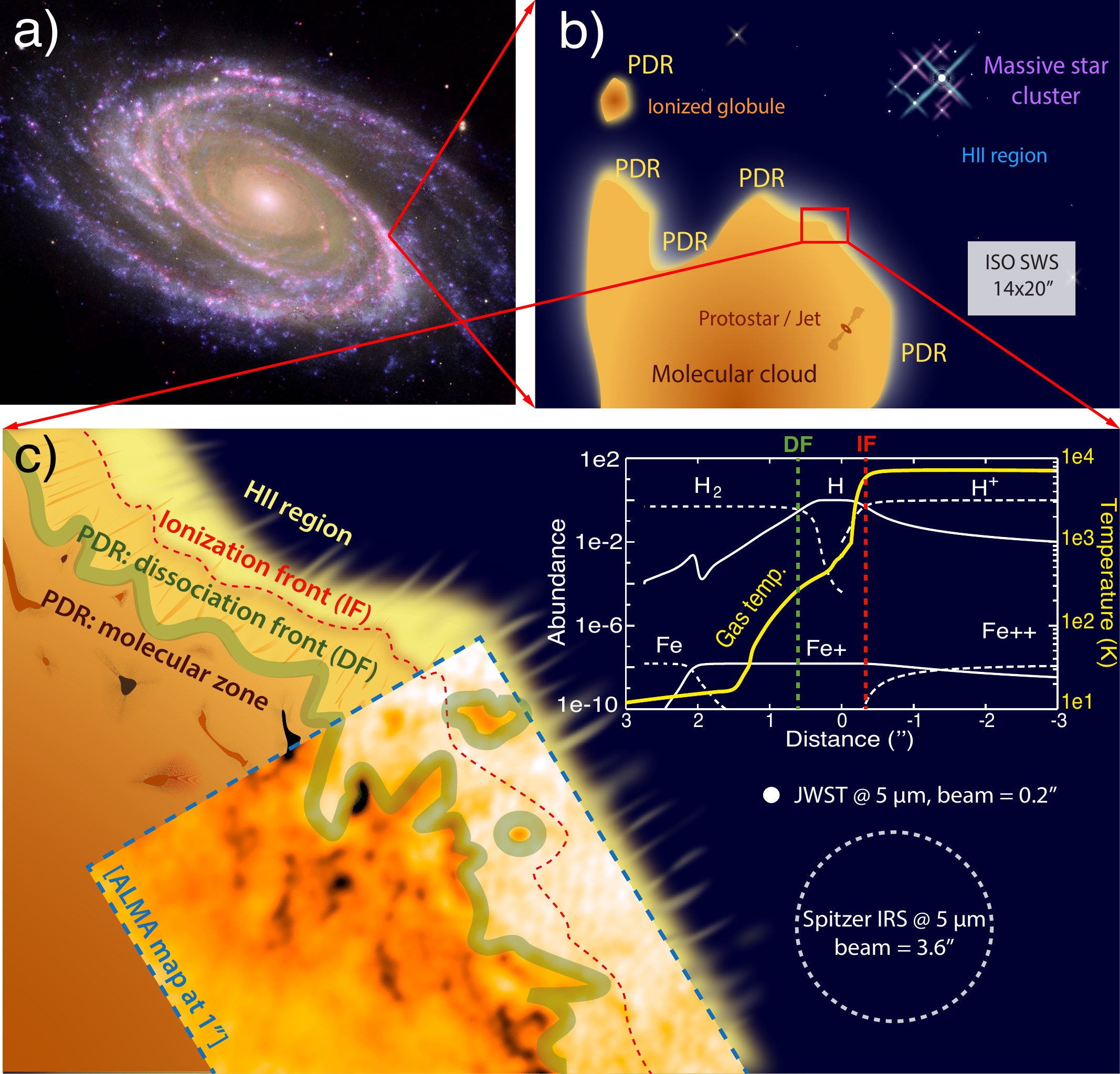}
    \caption{Zooming into a PDR. {\bf a)}~Multi-wavelength view of a galaxy (M81): UV-tracing massive stars (blue), optical-light-tracing \HII\, regions (green), and emission in the Aromatic Infrared Bands (AIBs) tracing PDRs (red).  Credits: Hubble data: NASA, ESA, and A. Zezas (Harvard-Smithsonian Center for Astrophysics); GALEX data: NASA, JPL-Caltech, GALEX Team, J. Huchra et al. (Harvard-Smithsonian Center for Astrophysics); Spitzer data: NASA/JPL/Caltech/S. Willner (Harvard-Smithsonian Center for Astrophysics). {\bf b)} Sketch of a typical massive star-forming region. {\bf c)}~Zoom in on the PDR, showing the complex transition from the molecular cloud to the PDR dissociation front, the ionization front and the ionized gas flow. 
    Inserted is the ALMA molecular gas data of the Orion Bar, at a resolution of 1$\arcsec$ \cite[dashed lines;][]{goicoechea}. The plot shows a model of the structure of the PDR. The scale length for FUV photon penetration corresponds to a few arcsec. The beam sizes of ISO-SWS, Spitzer-IRS, and JWST-MIRI are indicated. JWST will resolve the 4 key regions: the molecular zone,  the H$_2$ dissociation front,  the ionization front, and the fully ionized flow into the \HII{} region (see Section~\ref{sec:PDRs}). }
    \label{fig:pdr}    
\end{figure*}

\subsection{Photo-dissociation regions}
\label{sec:PDRs}

Photo-Dissociation Regions (PDRs; Fig.~\ref{fig:pdr}) are regions near massive stars, where most of the gas is neutral (i.e. H or H$_2$) and is heated by far-ultraviolet (FUV) photons (i.e. $6$~eV~$<E<13.6$~eV). 
Four key zones can be identified across the PDR: the {\sl molecular zone}, where the gas is nearly fully molecular, dense and cold (several tens of K); the {\sl H$_2$ 
dissociation front (DF)}, where the gas converts from nearly fully molecular to atomic and the temperature rises from 30 to 300 K; the {\sl ionization front (IF)} where the gas converts from fully neutral to fully ionized and the temperature increases to 7000\,K; and the fully ionized flow into the {\sl \HII~region}.
PDRs have a gas density ($n_{\rm H}$ for the hydrogen nucleus density) ranging from $n_{\rm H}\sim$10$^3$ cm$^{-3}$ in diffuse gas to $n_{\rm H}\sim$10$^6$ cm$^{-3}$ in dense star forming regions. The incident flux of the FUV field on the PDRs, $G_0$, is commonly characterized in units of the Habing field corresponding to $1.6 \times 10^{-3}$ erg\,s$^{-1}$\,cm$^{-2}$ when integrated between 6 and 13.6 eV \citep[][]{Habing68}. The standard interstellar radiation field has a $G_0$ value of $\sim$1.7 \citep{draine1978,Parravano03} while in most PDRs, $G_0$ is higher and can go up to a few $10^5$. The kinetic temperature of the gas ($T_g$) in PDRs lies in between ten and a few thousands degrees. Large dust grains (0.1 $\mu$m and above) are in equilibrium with the radiation field at temperatures of $\sim$30-70~K \citep[e.g.][]{Anderson:12, Paladini:12}. 
Highly irradiated PDRs can be found surrounding young \HII\, regions formed by O and early B stars \citep{stenberg}. 
At the ionized front, the gas is compressed by expansion of the ionized gas, the FUV radiation field is high ($G_0=10^{4-5}$), and densities and temperatures are also high ($n_{\rm H}\sim10^{5-6}$ cm$^{-3}$ and $T_g$ $\sim$ 100 to 2000 K). These are the physical conditions found in one of the most studied PDRs, the Orion Bar  \citep[see e.g.][]{tielens:85b, goicoechea}. Around less massive B stars, or at the interfaces of more evolved \HII\ regions, PDRs have a lower FUV radiation field ($G_0=10^{2-4}$) as observed for instance in PDRs like NGC~2023 \citep{Burton:98, Sheffer11}, NGC~7023 \citep{Fuente03}, or the Horsehead nebula \citep{Habart2011,Pabst17}. The diffuse medium in the Milky Way is a vast, low density ($n_{\rm H}\sim 10$ cm$^{-3}$) and low FUV radiation field ($G_0=1-10$) PDR \citep{vandishoeck86}, where the temperature is of the order of 50-100 K \citep{wolfire_neutral_2003}. 

Inside PDRs, the gas is mainly heated by the photo-electric effect 
working on Polycyclic Aromatic Hydrocarbon (PAH) molecules and small dust grains \citep{Verstraete1990, bakesandtielens94, weingartner_photoelectric_2001}.
Deep in the PDR, where the FUV radiation field is attenuated due to dust absorption in the upper layers, collisions of the gas with warm dust grains and cosmic-ray heating become important \citep[e.g.,][]{Tielens_book05}. Other heating mechanisms through H$_2$ formation or UV pumping of H$_2$ followed by de-excitation may provide additional important heating sources near the edge of dense PDRs \citep[e.g.,][]{lepetit,lebourlot2012}. The gas in PDRs is mostly cooled by far-IR (FIR) fine-structure lines, such as the [CII] line at 158 $\mu$m and the [OI] lines at 63 and 146 $\mu$m \citep[e.g.,][]{Hollenbach99,Bernard-Salas12}.
The interplay between heating and cooling mechanisms results
 in a large temperature gradient from the ionized to molecular gas across the PDR (Fig.~\ref{fig:pdr}c). 

Models have been very successful in explaining large-scale Galactic and extra-galactic observations of PDRs \citep[e.g.][]{tielens:85, sternberg, abgrall,wolfire90,lebourlot, 2006ApJ...644..283K, comparison07, cubick08}. However, Atacama Large Millimeter/submillimeter Array (ALMA) observations unambiguously revealed a highly sculpted interface between the molecular clouds and the ionized gas \citep[Fig.~\ref{fig:pdr};][]{goicoechea} and have challenged the traditional ``layered structure'' view of PDRs (and their models). 
Moreover, recent near-IR images obtained with the Gemini or Keck telescopes at high spatial resolution ($\sim$0.1$\arcsec$), similar to JWST, revealed with a spectacular level of detail structures unexpected within the classic irradiated molecular cloud \citep[e.g.][]{Hartigan2020,Habart2022}.
Series of ridges that follow along the interfaces may be associated with a multitude of small dense highly irradiated PDRs. JWST will resolve and observe directly the response of the gas to the penetrating FUV photons and give for the first time insight into the physical conditions and chemical composition of this very structured irradiated medium. 

\subsection{PDRs everywhere}
\label{sec:PDRseverywhere}

Stars in galaxies only form in cold gas, hence the efficiency at which the energy of FUV photons from massive stars is transferred to the interstellar gas in PDRs has a critical impact on the star-formation rate. This efficiency can typically be monitored using the mid-IR emission of nearby and distant galaxies \citep[e.g.][]{hel01,Peeters:pahtracer:04, Maragkoudakis:18, mck20, Calzetti:20}. Mid-IR observations are also useful to disentangle the contribution of shock versus PDR gas heating in galaxies (e.g. merger versus starburst \citealt{vandenAncker:00, guillard2012strong}). Dense ($n_{\rm H}>10^4$~cm$^{-3}$) and highly irradiated ($G_0\sim10^{4-6}$) PDRs are also present in the FUV-illuminated surfaces of protoplanetary disks \citep[e.g.,][]{vis07,woitke2009radiation} and govern the mass loss of these objects through photo-evaporation \citep{gorti2009time} in competition with planet formation.
IR observations give direct diagnostics of the FUV energy injected into the gas \citep[e.g.,][]{Meeus2013} and, in combination with PDR models, can constrain the physical parameters ($T_{g}$, $n_{\rm H}$) inside the FUV-driven winds of these disks \citep[e.g.,][]{Champion17}. 
Observations and models of planetary nebulae show that a large fraction of the gas ejected by evolved stars goes through a PDR phase \citep{hollenbach1995time, bernard2005physical} before being injected in the interstellar medium (ISM). Here again, IR spectroscopy provides key information on the initial physical and chemical properties of material in this phase \citep[e.g.,][]{Bernard-Salas:09, cox2015polycyclic} and allows the photo-chemical evolution of molecules, nanoparticles, and grains to be probed. 
Clearly, PDRs are present in a wide variety of environments. 

\subsection{IR signatures of PDRs}
PDRs emit mainly in the IR. Key PDR signatures in the near- and mid-IR (1-28~$\mu$m) include i) the Aromatic Infrared Bands (AIBs, Fig.~\ref{fig:OMC1}), 
attributed to Polycyclic Aromatic Hydrocarbons (PAHs) and related species that are heated by absorption of UV photons
\citep{leger_puget84, allamandola_polycyclic_1985}, ii) continuum emission, attributed to very small carbonaceous grains \citep[VSGs,][]{desert_interstellar_1990}, iii) ro-vibrational and pure rotational lines of H$_2$ \citep{Rosenthal2000},  and iv) emission lines from atomic ions \citep[e.g. S$^+$, Si$^+$, Fe$^+$, ][]{2006ApJ...644..283K}, and v), in regions close to massive stars, emission  from larger grains composed of silicates \citep{draine_interstellar_2003} or a mixture of silicates and carbonaceous material  at thermal equilibrium \citep{draine_infrared_2007, Jones2013}. \\

A large fraction of the observations that will be obtained with JWST will correspond to emission that is forged inside PDRs. It is therefore of paramount importance to understand how the observed mid-IR emission fingerprints are linked with physical conditions, and how these observations can be turned into probes of astrophysical environments. This requires a detailed knowledge of how PDRs ``work," by observing nearby, extended PDRs. This ERS program is designed to characterize the IR signatures of the Orion Bar PDR, to unravel the underlying physical and chemical processes, and validate diagnostic tools in order to facilitate the interpretation of PDR emission to be seen with JWST, in the local and distant Universe.




\vspace{2cm}

\section{Immediate goals of this ERS program}
\label{sec:immediate-goals}
The main goal of our program is to rapidly deliver ``template data", as well as data processing and analysis tools for PDRs, which will be crucial for JWST proposal preparation for both Galactic and extra-galactic sciences communities. To reach this ambitious goal, we have identified the following three immediate objectives: 
\begin{enumerate}
    \item Characterize the {\it full} \mbox{1-28 $\mu$m} emission spectrum of the key zones and sub-regions within the ionized gas, the PDR, and the surrounding molecular cloud and determine the physical and chemical conditions in these specific environments. 
    \item Examine the efficacy and limitations of the narrow/broad band filters in the study of PDRs by accurately calibrating narrow/broad band filters which capture gas lines and AIBs as PDR probes.
    \item Deliver tools facilitating post-pipeline data reduction and processing, as well as PDR and AIB analysis tools required for the interpretation of (un)resolved PDR emission to be observed with JWST. 
\end{enumerate}

\begin{figure*}[t]
\begin{center}
\includegraphics[clip,trim =0.cm 2.3cm 0.cm 3.5cm,scale=0.32]{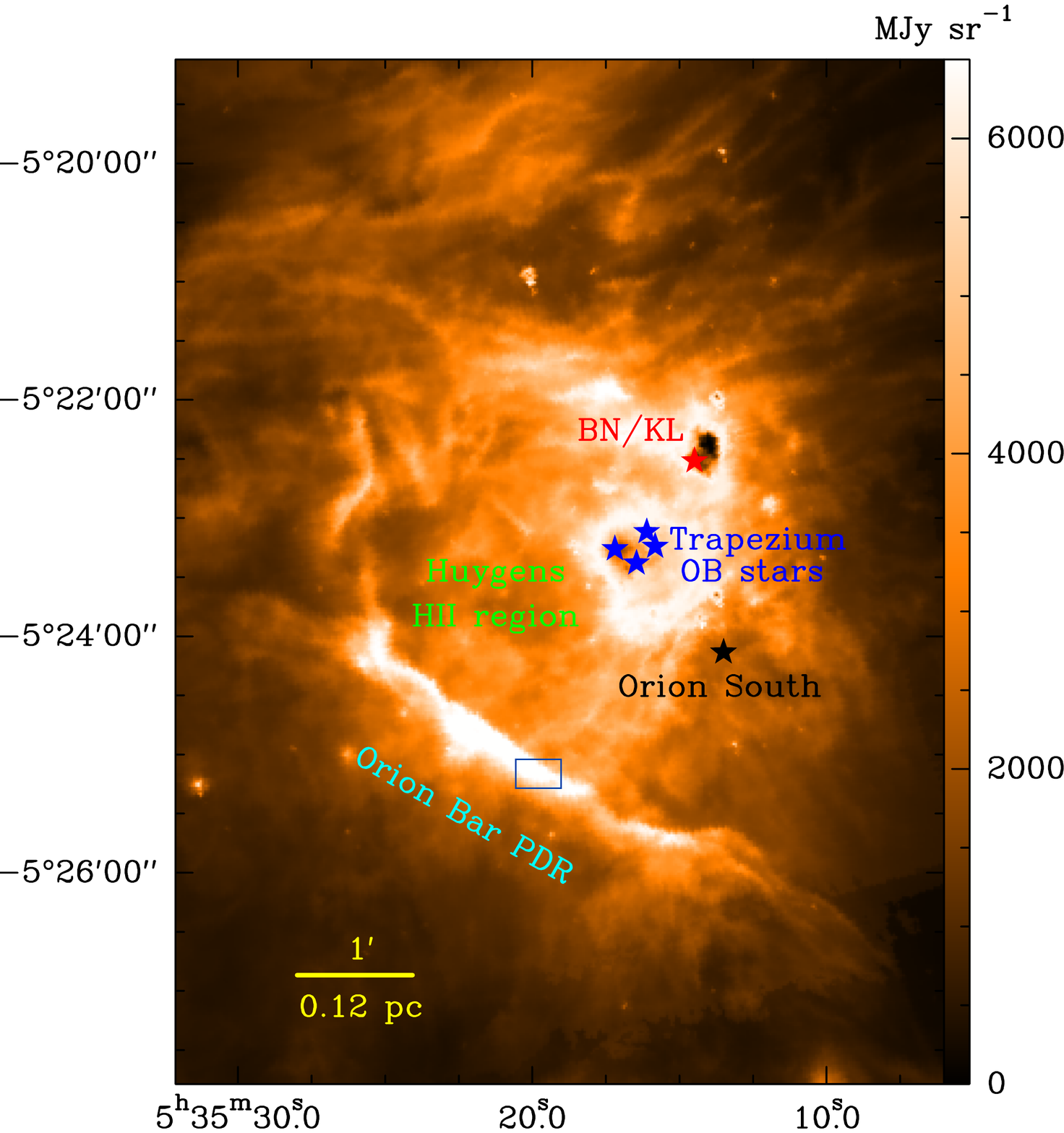}
\includegraphics[clip,trim =0.cm 0.5cm 0.cm 1cm,scale = 0.3]{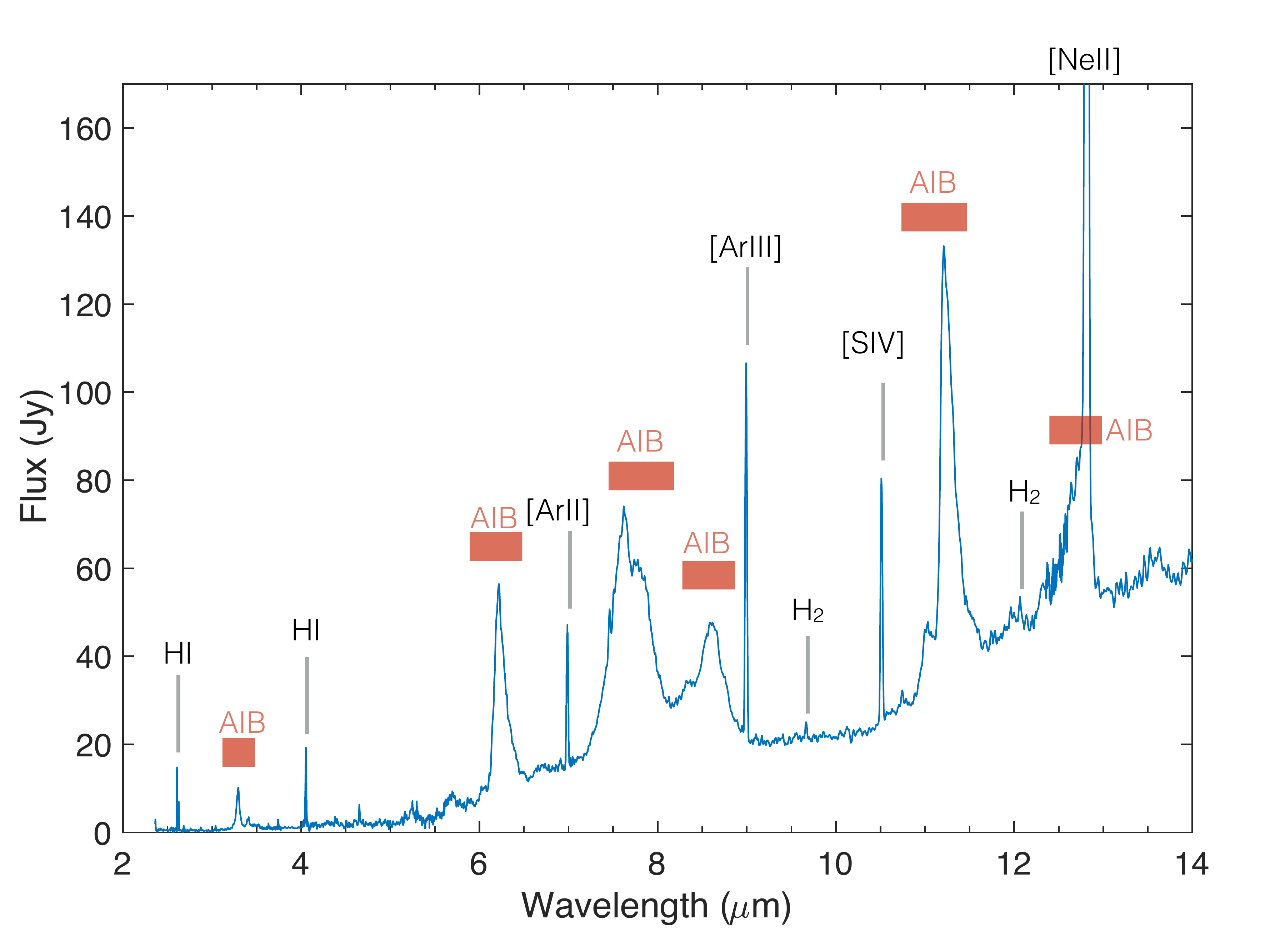}
\end{center}
\caption{The Orion Bar in the mid-IR. {\it Left:} Overview of the central region of the Orion Nebula as seen with 
{\emph Spitzer}-IRAC at 8$\mu$m. Salient objects, including the Orion Bar which is the target of this program, are labeled. Figure adapted from \citet{goicoechea2015}. {\it Right:} ISO-SWS spectrum of the Orion Bar PDR extracted at the position indicated by the blue box in the image (for the 2.3-12 $\mu$m range). Main spectroscopic fingerprints are labeled. 
}
\label{fig:OMC1}
\end{figure*}

To this end, we will observe the Orion Bar, a proto-typical PDR situated in the Orion Nebula (Fig.~\ref{fig:OMC1}) using NIRCam, NIRSpec, and MIRI in the \mbox{1-28 $\mu$m} wavelength range. The Orion Bar has a well-defined UV illumination and geometry making it an ideal target to reach our science goals. These observations will, for the first time, spatially resolve and perform a tomography of the PDR, revealing the individual IR spectral signatures in the four key zones of a PDR: the molecular zone, the H$_2$ dissociation front, the ionization front, and the ionized flow into the \HII~region (Fig.\,\ref{fig:pdr}). 
%
%

\section{Target: the Orion Bar}
\label{sec:target}

\begin{figure*}
\begin{center}
\includegraphics[scale=0.45]{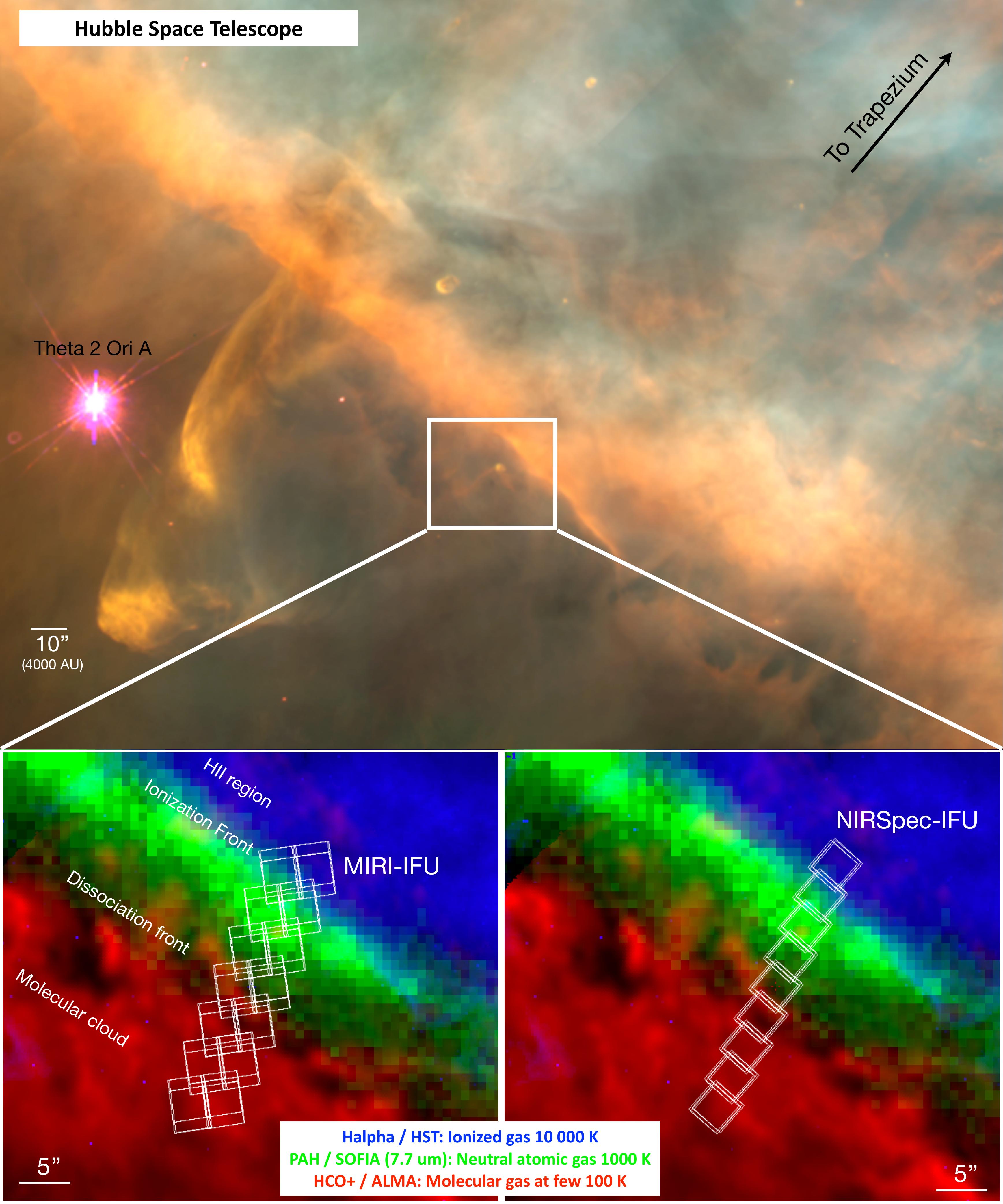}  
\end{center}
\vspace*{0cm}
\caption{Overview of the Orion Bar region at visible wavelengths, observed with the Hubble Space Telescope (top panel; credits: NASA, C.R. O'Dell and S.K. Wong, Rice University). Lower panels zoom on the region of interest showing the footprints of the MIRI-IFU and NIRSpec-IFU mosaics on a multi-wavelength view of the Orion Bar composed of H$\alpha$ 656~nm emission \citep[blue,][]{bally2000}, PAH 7.7 $\mu$m emission \citep[green,][]{Salgado16}, and HCO$^+$ (4-3) 356.7 GHz emission \citep[red,][]{goicoechea}.}
\label{fig:overlay_spectrosopy}
\end{figure*}

Orion Bar (Fig.~\ref{fig:OMC1}) is a prototype of a strongly UV-irradiated PDR with a nearly edge-on geometry \citep[e.g.,][]{Hoger95}, convenient to  study and spatially resolve the structure, physical conditions, and chemical stratification of a PDR.
The Orion Bar is a bright (at many wavelengths, Fig.~\ref{fig:overlay_spectrosopy}) escarpment of the Orion molecular cloud (OMC), the closest\footnote{The most commonly adopted distance to the Bar is 414\,pc \citep{Menten07} although more recent observations, including Gaia, point to slightly lower values \citep{Kounkel17,Gross18}.} site of ongoing massive star-formation.  The Orion Bar is illuminated by the \mbox{O7-type} star \mbox{$\theta^1$ Ori C}, the most massive member of the Trapezium young stellar cluster, which lies at the heart of the Orion Nebula \citep[about 2$'$ north west of the Bar, e.g.,][]{Odell01}. The intense ionizing radiation 
and strong winds from \mbox{$\theta^1$ Ori C} power and shape the nebula \citep[][]{Gudel08,Pabst19}
The Bar (also referred to as the Bright Bar \citealt[e.g.,][]{Fazio74,Balick74,Werner76}) historically refers to the elongated rim near the ionization front (originally detected in the radio continuum and in optical lines) that separates the neutral cloud from the ionized \HII~gas with an electron temperature $T_{\rm e}$\,$\approx$10$^4$\,K and electron density $n_{\rm e}$ of several 1000\,cm$^{-3}$ \citep[e.g.,][and references therein]{Weilbacher15}. The UV radiation field incident on the Orion Bar PDR is G$_0= 1-4\times10^4$ \citep[e.g.,][]{Marconi98}. 
Beyond the ionization front, only FUV photons with energies below 13.6\,eV penetrate the cloud. 
The first PDR layers are predominantly neutral and \textit{atomic}: \mbox{[H]\,$>$\,[H$_2$]\,$\gg$\,[H$^+$]}. A plethora of NIR lines are emitted from this region \citep[e.g., CI recombination lines, OI fluorescent lines, see][]{Walmsley00}.  This warm and moderately dense  gas ($n_{\rm H}$ of a few 10$^4$\,cm$^{-3}$) is mainly heated by photoelectrons ejected from PAHs and grains and it is mainly cooled by the FIR [\CII]\,158\,$\mu$m and [\OI]\,63\,$\mu$m and 145\,$\mu$m fine-structure lines \citep[e.g.,][]{Tielens93,Herrmann97,Bernard-Salas12,Ossenkopf13}. The observed narrow (\mbox{$\Delta v$\,=\,2--3\,km\,s$^{-1}$}) carbon and sulfur radio recombination lines also arise from these  layers and provide a measure of the electron density in the PDR \citep[$n_{\rm e}$\,$\simeq$\,10--100\,cm$^{-3}$; e.g.,][]{Wyrowski97,Cuadrado19,Goicoechea21}. The atomic PDR zone also hosts the peak of the MIR AIBs  \citep[e.g.,][]{Bregman89,Sellgren90,Tielens93,Giard94,Knight21}.  

At about 15 $\arcsec$ from the ionization front (at \mbox{$A_V$\,$\simeq$\,1-2~mag} 
of visual extinction into the neutral cloud), the FUV dissociating photons are sufficiently attenuated and most of the hydrogen becomes molecular, H$_2$ molecules contains
over 90\% of the H nuclei \citep{vanderWerf13}. This  H/H$_2$ transition (dissociation front) displays a forest of near- and mid-IR rotational and vibrationally excited H$_2$ lines \citep[e.g.,][]{Parmar91,Luhman94,vanderWerf96,Allers05,Shaw09,Zhang21}, including \mbox{FUV-pumped} vibrational levels up to $v$\,=\,10 \citep{Kaplan17} and HD  rotational lines  \citep{Wright99,Joblin18}. Analysis of the IR H$_2$ and 21\,cm \HI~lines toward the dissociation front suggests warm kinetic temperatures for the molecular gas (\mbox{$T_{\rm g}$\,=\,400-700\,K}) which are not easy to reproduce by PDR models using standard (diffuse ISM) grain properties and heating rates \citep[e.g.][]{Allers05}.

 Beyond the dissociation front, between \mbox{$A_V$\,=\,1-2} and 4\,mag, the C$^+$/C/CO transition takes place \citep[e.g.,][]{Tauber95} and the PDR becomes \textit{molecular}, with the abundance of reactive molecular ions quickly rising \citep[e.g.,][]{Stoerzer95,Fuente03,Nagy13,Tak13}. 
 FIR and MIR photometric images  reveal that the thermal dust emission peaks deeper into the PDR than the AIBs. They show a dust temperature gradient from \mbox{$T_{\rm d}$\,$\simeq$\,70\,K} near the ionization front to \mbox{$T_{\rm d}$\,$\simeq$\,35\,K} in the less exposed layers \citep{Arab12,Salgado16}. Dust models of the Orion Bar  require FUV and IR grain opacities lower than in the diffuse ISM, with  \mbox{$R_V$\,=\,$A_V$/$E_{B-V}$\,$\simeq$\,5.5}, consistent with a flatter extinction curve   \citep[e.g.,][]{Cardelli89,Abel06}, and with larger-than-standard-size grains.

As the FUV flux drops, $T_{\rm g}$ and $T_{\rm d}$  decrease too. The intermediate $A_V$ layers of the Orion Bar PDR show a rich chemical composition which includes  a large variety of small hydrocarbons, 
complex organic species, and some deuterated molecules \citep[e.g.,][]{Hoger95,Simon97,Peeters04,Leurini06,onaka2014, Cuadrado15,doney2016, Cuadrado17}. At greater $A_V$, as $T_{\rm d}$ decreases, abundant elements such as oxygen and sulfur atoms start to deplete on dust grains and ices form. The ice mantle composition and the chemistry that takes place on PDR grain surfaces are still uncertain  \citep[e.g.,][]{Guzman11,Esplugues16}. Photodesorption of these ices enriches the gas with new chemical species \citep[e.g.,][and references therein]{Putaud19,Goicoechea21b}.  

Unfortunately, most of the observations described above (especially at  FIR and longer wavelengths) refer to modest angular resolution observations (10\arcsec--20$\arcsec$) that do not spatially resolve the main transition zones of the PDR. As a consequence, their fundamental structures: homogenous versus clumpy, constant density versus constant pressure, and the role of magnetic pressures and dynamical effects (e.g., photoevaporation or low-velocity shocks) are still debated.  This ongoing discussion has led to the development of very detailed PDR models  \citep[e.,g.][]{tielens:85,Burton90, Stoerzer98, Pellegrini09, Andree17, bron18, Kirsanova19}.  One of the controversial points is the need to invoke the presence of high density clumps (\mbox{$n_{\rm H}$\,=\,10$^6$--10$^7$cm$^{-3}$}) to explain the emission from  certain high critical density tracers \citep[e.g.,][]{Tauber94,vanderWerf96,YoungOwl00,Lis03}. Interestingly, the most massive clumps might collapse and form low-mass stars \citep[e.g.,][]{Lis03}. However, very small dense clumps may not exist, at least close to the dissociation front \citep[e.g.,][]{Gorti02}. Alternatively, some of the observed features  may be explained without invoking clumps by a roughly isobaric PDR, at high thermal-pressure (\mbox{$P_{\rm th}/k$\,$\approx$\,10$^8$\,cm$^{-3}$\,K}) \citep[e.g.,][]{Allers05,Joblin18} embedded in a more diffuse medium.  In addition, the magnetic pressure may play a role in the PDR/clump dynamics \citep[e.g.,][]{Pellegrini09,Pabst20}. Recent observations of the FIR dust polarization suggest a 
plane-of-the-sky magnetic field strength of $B_0$\,$\simeq$\,300\,$\mu$G \citep{Chuss19,Guerra21}.

 Recently, ALMA observed the same FOV that JWST will target, providing  \mbox{$\sim$\,1\,$\arcsec$} resolution images of the CO and HCO$^+$ emission \citep{goicoechea}. Instead of a homogenous PDR with well-defined and spatially separated \mbox{H/H$_2$} and \mbox{C$^+$/C/CO} transition zones, ALMA revealed rich small-scale structures (akin to filaments and globules), sharp edges, and uncovered the presence of an embedded proplyd \citep[object \mbox{203-506};][]{Champion17}. The CO gas temperature just beyond the dissociation front is $T_{\rm k}$\,$\simeq$\,200\,K \citep[see also][]{Habart10,Joblin18} and decreases deeper inside. 
  
\section{Scientific objectives}
\label{sec:science-objectives}

The unprecedented dataset will obtain with Webb (seeSect.~\ref{sec:obs}) will allow us 
to address several science questions. In this section we highlight three science objectives that can be tackled with this ERS program.  \\

\begin{figure*}[t]
{\includegraphics[angle=0, scale=0.64]{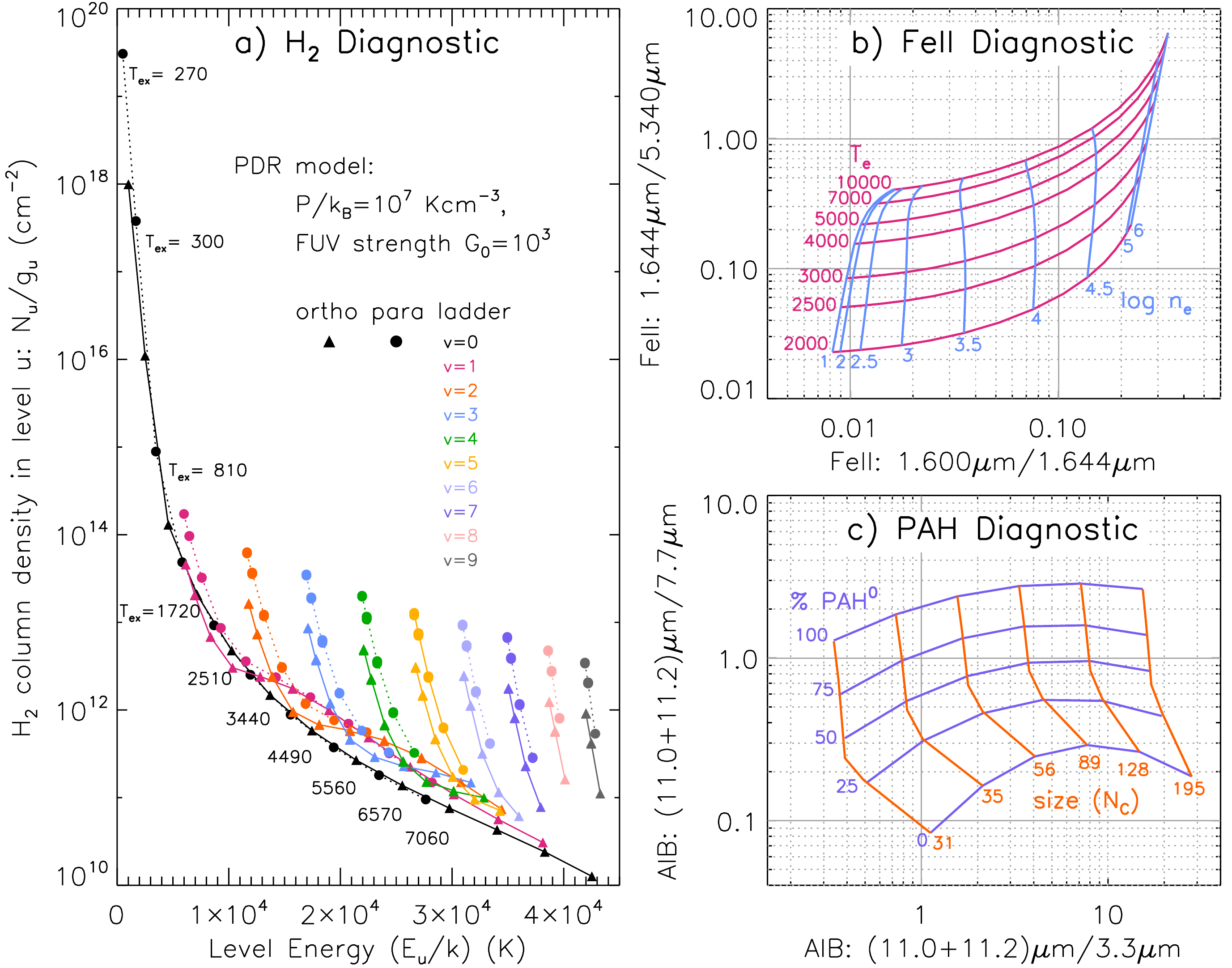}}
\caption{Example IR PDR diagnostics: {\bf a)} Excitation diagram from ${\rm H_2}$ lines in PDRs, that are observable by MIRI and NIRSpec, as a tracer of the warm and hot (UV-pumped) excitation temperatures (excitation temperatures derived from the level populations by a local Boltzmann fit are indicated); {\bf b)} The [\FeII{}] lines as a tracer of the temperature and density distribution from the ionized gas to the PDR \citep{chianti}; {\bf c)} AIB emission ratios as a tracer of PAH size (in function of number of carbon atoms, N$_C$) and charge (in function of the neutral PAHs fraction, i.e. PAH$^0$/(PAH$^0$+PAH$^1$)), computed from harmonic IR spectra using PAHdb and assuming an interstellar radiation field (see Section~\ref{sec:SEPs} for a description of PAHdb). Figure adapted from \citet{Maragkoudakis:20}.}
\label{fig:diagnostics}
\end{figure*}

\paragraph{\textbf{Evolution of the physical and chemical conditions at the critical H$^+$/H$^0$/H$_2$ transition zones}}
We expect to detect and spatially resolve a large number of lines (see Section~\ref{sec:simulated-spectra} and Fig. \ref{fig:spectra}), i.e. fine-structure lines of several ions and atoms (e.g., [\FeII{}], [\FeI{}], [\ArIII{}], [\ArII{}], [\SIV{}], [\SII{}], [\SI{}], [\PIII{}], [\NeIII{}], [\NeII{}], [\NiII{}], [\FI{}], [\ClI{}], ...), fluorescent lines (O, N), recombination lines (H, He, C), pure rotational and ro-vibrational transitions of H$_2$ and its isotopologue HD (both collisionaly excited or radiatively pumped), ro-vibrational transitions of non-polar molecules 
(CH$_4$, C$_2$H$_2$, CO$_2$, C$_6$H$_6$), and possibly, for the first time in a PDR, highly excited pure rotational and ro-vibrational transitions of CO, H$_2$O (HDO), and OH. Observations of these species, which each provide a diagnostic of a specific physical environment or chemical reaction, at unprecedented high spatial resolution (up to 0.1$\arcsec$ or 40 AU at 400 pc)  for a PDR have so far been out of reach. 

Variations in the physical conditions and the high density sub-structures in the critical H$^+$/H$^0$/H$_2$ transition zones are poorly known, yet they are of fundamental importance for PDR models and data interpretation. The measurement of a large number of fine-structure lines of ions and atoms will give access to the warm plasma cooling and pressure gradients before the ionization front and between the ionization front and dissociation front \citep{osterbrock2006astrophysics}. Strong constraints can then be placed on metallicities,  electron densities and temperature variations (e.g., Fig. \ref{fig:diagnostics}b and \citealt{verma2003mid}). Benchmarking these probes of ionized / neutral gas interface is particularly important to support extragalactic studies (e.g. \citealt{cormier2012nature}).  
A description of the impinging UV radiation field (intensity and wavelength dependence) can also be obtained via fluorescent lines \citep[e.g.,][]{Walmsley00}.  A determination of the physical conditions in the neutral layer beyond the ionization front can also be assessed with recombination lines \citep[e.g.,][]{Natta94,Cuadrado19}. On the other hand, pure rotational and ro-vibrational lines of H$_2$ and possibly HD will provide a great thermometer for the bulk of the gas and pressure  gradients inside the PDRs \citep[e.g., Fig. \ref{fig:diagnostics}a and][]{Parmar91,Wright99,habart05,Allers05,Habart2011,Sheffer11, Kaplan17,Joblin18}.
 
These constraints on the physical conditions will be essential to study the dynamical effects in PDRs, e.g. compression waves, photo-evaporative flows, ionization front and dissociation front instabilities. Very precise determination of the offset between the ionization front and the dissociation front will be obtained, as well as, how this offset is affected by the shapes of the evaporative flows \citep{Carlsten2018}. Moreover, JWST will probe the thin surface layers that are sufficiently heated
to photo-evaporate from the PDR. 

These observations will allow for better understanding of the physical and dynamical processes at work, identify pertinent signatures and diagnostic tracers for the different key PDR zones, improve model predictions for both warm molecular and ionized gas and help the development of new PDR models which couple the dynamics of photoevaporation to physico-chemical processes \citep[e.g.,][]{bron18}.


The role of dust properties (e.g., size distribution) in determining the position of the H$^+$/H$^0$/H$_2$ transition  will also be better constrained with these observations \citep[][]{Allers05,schirmer2021}. 
Moreover, grain surface chemistry is an unavoidable route for efficient H$_2$ formation \citep[e.g.,][]{gould1963, Wakelam2017}.
Observations of numerous H$_2$ rotational and ro-vibrational lines at high spatial resolution might constrain both the H$_2$ formation processes in warm gas and grains and the mechanisms that control the H$_2$ ortho-para ratio \citep[e.g.,][]{habart04,bron14,bron16}.
Determination of the H$_2$ formation rate on interstellar grains and its abundance is particularly important, as it controls most of the PDR physical structure and subsequent development of the chemical complexity in the ISM \citep[for a review on H$_2$ formation in the ISM see][]{Wakelam2017}. 

Possibly, highly excited rotational and  ro-vibrational lines of CO, H$_2$O, HDO, OH, CH$^+$   will also lead to a better understanding of the radiative and chemical pumping in PDRs. 
The presence of large columns of vibrationally excited H$_2$ that help to overcome certain reactions drive the endothermic carbon chemistry \citep[e.g.,][]{Sternberg95,Agundez10}. 
Mid-IR superthermal emission of OH will probe the far-UV dissociation of H$_2$O and measure the local irradiation and density conditions \citep[]{Tabone2021}. 
Finally, ro-vibrational lines of non-polar molecules (e.g., CH$_4$, C$_2$H$_2$, C$_6$H$_6$) will give a more complete inventory of hydrocarbon species and better characterize their formation/destruction processes via top-down or bottom-up chemistry \citep[e.g., ][]{Cernicharo_2004,Parker2012,Contreras:13, Pilleri2013,Alata2014,jones2015,guzman2015, Sciamma:20}.

\paragraph{\textbf{Photochemical evolution of carbonaceous species}}

A key spectroscopic feature of PDRs is the AIBs, observed throughout the Universe
at 3.3, 6.2, 7.7, 8.6, and 11.2 $\mu$m (see Fig.~\ref{fig:OMC1}) and attributed to the
infrared fluorescence of nanometric particles and molecules from the family of Polycyclic Aromatic Hydrocarbons (PAHs) 
\citep{leger_puget84, allamandola_polycyclic_1985}.
In addition to interstellar PAHs, emission from the fullerene C$_{60}$ is also present in PDRs, characterized by emission bands detected at 7.0, 17.4 and 18.9 $\mu$m \citep{sellgren2010c60, Peeters:12, boe12, castellanos_c_2014_fc, berne_detection_2017}. The underlying continuum present in mid-IR spectra (Fig.~\ref{fig:OMC1}) is more difficult to attribute, but is believed to be due to some form of very small carbonaceous grains (VSGs, \citealt{desert,compiegne}), amorphous hydrocarbon nano-particles \citep{Jones2013} and/or PAH clusters \citep{Rapacioli2006}. 
In regions closest to massive stars, large silicate grains can also emit in the mid-IR continuum \citep[][]{Cesarsky2000, Knight21:Orion}. 

An important aspect of these spectroscopic features is that their relative contributions to the mid-IR spectrum vary significantly (e.g. \citealt{Joblin:3umvsmethyl:96,  Cesarsky1996,Sloan:97, Peeters:prof6:02, rapacioli2005,compiegne_aromatic_2007,Povich:07, Watson:08, boe12,Candian:12,Mori:12, Mori:14, Stock:16, Sidhu:21}). These variations suggest an evolution from VSGs with a possible mixed aromatic aliphatic nature present in UV shielded regions, to free flying PAH species at the surface of molecular clouds \citep{rapacioli2005, berne_blind_2012, pilleri2015,peeters17, murga2020,schirmer2020}, and eventually to more stable fullerenes \citep{berne2015} and GrandPAHs \citep[i.e. the most stable PAHs;][]{andrews} in harsh (high $G_0$) environments. 
This evolution in PDRs has strong implications for the understanding of the PDR physics, notably the local extinction of the UV field, the heating of the gas (by photoelectric effect), and the formation of H$_2$. Without detailed knowledge of the properties of these species, the implementation of these mechanisms in theoretical models can only be approximate.
In addition, this general scenario needs however to be extended, by identifying the connection between the photochemical evolution of large carbonaceous species such as PAHs and fullerenes with other chemical networks of PDRs. One critical aspect concerns the link with small hydrocarbons such as acetylene or benzene. These non polar species cannot be detected with radio-telescope, but their infrared band may be detected with JWST.  Mapping the emission from these species in the Orion Bar at high angular resolution will offer an unprecedented access to the organic inventory of a PDR, allowing to link chemical networks. 

\begin{figure*}[t]
    \centering
    \includegraphics[scale=0.45]{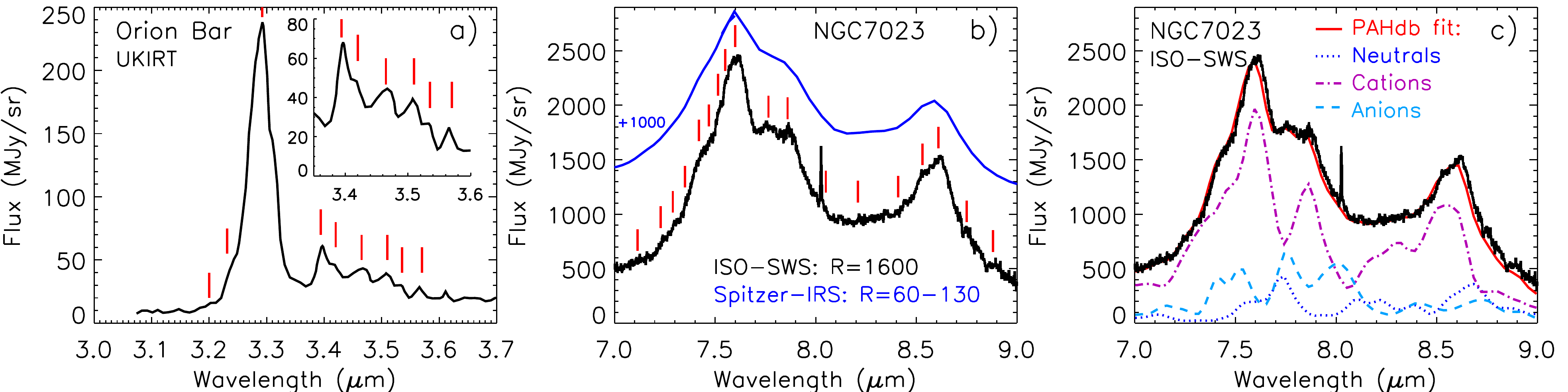}
    \vspace*{-.25cm}
    \caption{The spectral richness of the AIB emission toward two prototypical PDRs shown for {\bf a)} the 3 $\mu$m and {\bf b)} the 8 $\mu$m region \citep{geballe, moutou}. The inset in panel a zooms in on the 3.4-3.6 $\mu$m region. Vertical bars indicate sub-structure (reflecting sub components). These are not detectable at low spectral resolution (panel {\bf b}, blue line, offset=1000). {\bf c)} PAHdb fitting to the 7--9 $\mu$m range shown with its breakdown in charge states (see Section~\ref{sec:SEPs} for a description of PAHdb).  }
    \label{fig:PAHs}    
\end{figure*}

The combination of high spectral resolution mid-IR spectroscopy of AIBs (resolving the plethora of AIB subcomponents and structure; see Fig.~\ref{fig:PAHs}) with detailed knowledge of the physical conditions will provide key constraints on astrochemical models of PAH evolution \citep[e.g.][]{Galliano:08, montillaud2013, Mori:14, andrews, berne2015, boersma:15, croiset, Stock2017, murga2019,  dartois20, Knight21:Orion, Knight21,murga2022}, which will help determine the dominant properties of PAH populations. This will in turn guide laboratory experiments or quantum chemical calculations which provide the molecular parameters these models rely on \citep[e.g.,][]{Maltseva:15, Sabbah:17, Salama:18, Martinez:20, dartois20, Wiersma:20, Gatchell:21, Zettergen:21}. Overall, this synergy between observations, models, laboratory experiments and theory is essential to determine the properties and role of PAHs and related carbonaceous species in space (see \citealt{joblin2011pahs} and references therein). 

Finally, the comparison of JWST observations with dust emission models will also be essential to constrain the dust evolution on a small scale at the edge of the PDR. Earlier studies using suggest that the abundance of VSGs is reduced in a variety of PDRs (e.g., in filaments of the Taurus cloud, the Horsehead and NGC 2023/7023 nebula,\citealt{stepnik2003,compiegne2008,Arab12}), but it is unclear what the origin of this depletion is.  

\paragraph{\textbf{Interpretation of unresolved PDRs}}

The JWST spectra of various types of objects (typically galaxies or UV-irradiated protoplanetary disks) will be dominated by PDR emission which is spatially unresolved. Deriving the physical conditions in these sources, where the emission from all PDR components is blended in one or a few spectra, is therefore much more difficult. This ERS program will provide Science-Enabling Products that will facilitate the interpretation of unresolved PDRs (Section~\ref{sec:SEPs}). 
For instance, we will provide template spectra for the four key regions of a PDR which can provide a first benchmark to interpret unresolved PDR emission. 
We will extend existing data-analysis tools and line and AIB diagnostic tools into the JWST era and validate them on the ERS data from this program. 
This approach has been highly successful in the past: tools which were benchmarked on galactic star-forming regions or nearby galaxies, such as PDR models (e.g. the Meudon code \citealt{lepetit}, the Kosma-Tau PDR code \citealt{markus}, the \citealt{kaufman99} model, the UCL-PDR model \citealt{bell2006molecular}) or spectroscopic decomposition tools (e.g. PAHFIT \citet{smith}, PAHTAT \citet{pilleri12}) have also been widely used for the analysis and the determination of physical conditions in external galaxies \citep[e.g.][]{malhotra2001far,chevance2016milestone,Naslim15, cormier2012nature, bayet2009molecular} or disks \citep{Champion17, berne_what_2009}.


\section{Simulated spectra}
\label{sec:simulated-spectra}

\begin{figure*}[t]
    \centering
    \includegraphics[width=\linewidth]{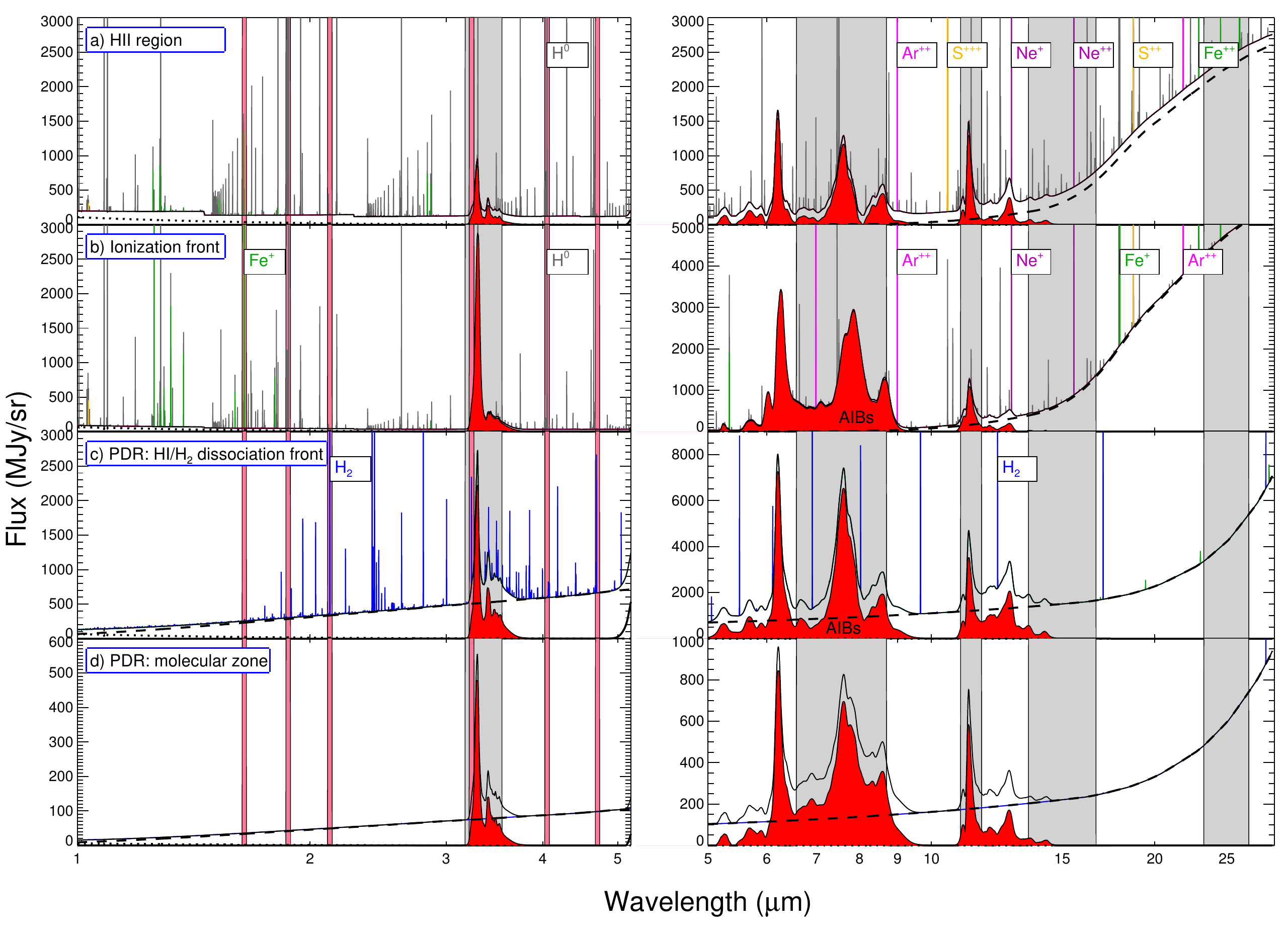}
    \caption{Model IR spectra at a spectral resolving power of 3000 of the 4 key regions within 
    the interface of the \HII\, region around the massive stars and the molecular gas (Fig. 1) illustrating the spectral richness that JWST will observe. Dust-scattered light and continuum emission are shown in dotted  and dashed lines. Ionic, atomic, and molecular gas lines are shown in colors (grey, green, pink, purple, blue). Aromatic and small dust bands are shown in red.  The band-passes of the photometric filters selected in this ERS program are shown in gray and pink for the medium and narrow filters respectively. Spectra have been calculated 
    with the Cloudy \citep{cloudy}, Meudon PDR code \citep[][]{lepetit}, the spectra from  \citet{foschino2019} for the PAH emission, and the DustEM \citep{compiegne} and THEMIS/SOC models \citep{Schirmer2022} for the dust emission and scattering. 
    } 
    \label{fig:spectra}
\end{figure*}

In order to illustrate the spectral richness that JWST will observe and to perform signal-to-noise ratio (SNR) calculations, we modelled the IR emission of the 4 key regions between the interface of the \HII~region around the massive stars and the molecular gas (Figure~\ref{fig:pdr}), for the case of the Orion Bar (Sect.~\ref{sec:target}). These spectra are also used by our team to create simulations of the future JWST observations to test the Science-Enabling Products (Sect.~\ref{sec:SEPs}) and advanced data-processing algorithms \citep[e.g.,][]{guilloteau2020, guilloteau2020b}.  Figure~\ref{fig:spectra} shows the obtained model templates at a spectral resolving power of 3000 illustrating the contribution of ionic, atomic, and molecular gas lines, AIBs, small dust bands, dust scattered light, and continuum emission.  The spectra are available in numerical format from \red{[link to be inserted upon publication]}. The four spectra have been computed individually for each region and each component using (i) the Cloudy code for the ionized gas in the \HII~region and the ionization front \citep[][]{cloudy};
(ii) the Meudon PDR code for the contribution from the atomic and molecular lines in the neutral PDR gas \citep[][]{lepetit};  
(iii) the AIB spectra extracted by \citet{foschino2019};
and (iv) the DustEM tool and SOC radiative transfer code for the contribution from the dust continuum and scattered light \citep{compiegne,juvela2019,schirmer2021}.
Below we briefly describe the parameters and calculation requirements used for each model and region. The physical parameters used  for  these  models  correspond  to  those  of  the  Orion  Bar described in Section~\ref{sec:target}. 
\\
{\bf Ionized gas emission.} For the ionized gas component we rely on the Cloudy code \citep{cloudy}. We adopt an illuminating star 
characterized by an O star model with an effective temperature $T_{eff}$~=~40,000~K. The TLUSTY stellar model was used. The total number of ionizing photons emitted by the star is set to \mbox{$Q_{\rm LyC}$\,$=$\,7$\cdot$10$^{48}$~photon\,\,s$^{-1}$} \citep[similar to that used by][]{Pellegrini09}.
The separation between the center of the source and the illuminated face of the cloud is assumed to be 0.01~pc. For the density, the initial electronic density is assumed to be $n^0_e$~=~3000~cm$^{-3}$. A power-law density gradient irradiated from the outside is taken for the density profile assuming an exponent $\alpha$=2 and the cloud scale depth $R_{scale-depth}$=0.3 pc \citep[similar to that used by][]{Shaw09,Pellegrini09}.
\\
{\bf Molecular and neutral atomic gas emission.} For this component, we rely on the Meudon PDR code. 
We consider an isobaric model with a thermal pressure \mbox{$P=4\cdot 10^8$\,K\,cm$^{-3}$} based on \cite{Joblin18}. We fix the radiation field impinging on the PDR so that, at the edge of the PDR, $G_0\,=\,2.25\cdot10^4$ in Habing units \citep[in agreement with previous estimates given $G_0\,=\,1-4\cdot10^4$ ][]{tielens:85b,Marconi98}.
We adopted the extinction curve of HD~38087 of \citet{Fitzpatrick1990} and $R_V\,=\,5.62$ which is
close to the value determined for Orion Bar of 5.5 \citep{Marconi98}. 
A complete transfer with exact calculation of the mutual screening between the UV pumping lines of H$_2$ was done since it can have a significant effect on the position of the H/H$_2$ transition and it can impact the intensities of the ro-vibrational lines of the radiative cascade. 
To obtain the model template spectra, we calculate the cumulative line intensities from the atomic and H/H$_2$ transition region ($0<A_V<2.5$) and from the molecular region (which starts at the C/CO transition, $2.5<A_V<8.5$). We set an upper limit of $A_V$=8.5 to the molecular region to eliminate emission caused by the interstellar radiation field on the back side. 
This PDR model provides a good agreement with the observed values of the high-J CO emission produced before the C$^+$/C/CO transition and that originate from small structures of typical thickness of a few 10$^{-3}$\,pc or $\sim$1$\arcsec$ \citep{Joblin18}. 
To reproduce the nearly edge-on geometry of the Bar \citep{wen1995three,hogerheijde1995millimeter,jansen1995millimeter,walmsley2000structure},
we adopt a geometry in which the PDR is observed with a viewing angle $\theta$ between the line-of-sight and the normal to the PDR equal to $\sim$60$^{\circ}$. This angle is defined with 0$^{\circ}$ being face-on and 90$^{\circ}$ edge-on. The value of 60$^{\circ}$ gives an approximation of a nearly edge-on PDR and is the maximum inclination that can be used to derive line intensities in the 1D PDR Meudon code. 
The optically thin line surface brightnesses are enhanced by a geometrical factor of  1/cos($\theta$)=2 relative to their face-on surface brightnesses.
The uncertainty on this angle could lead to an additional scaling factor on all line intensities. 
For simplicity, we adopt the same geometrical factor for the PDR and Cloudy models. 
\\
{\bf Dust emission and scattering.} 
To compute the dust emission and scattering in the neutral zone, we use the THEMIS\footnote{The Heterogeneous dust Evolution Model for Interstellar Solids, available here: \url{https://www.ias.u-psud.fr/themis/}. See model overview in \citet{jones2017} and references therein.} interstellar dust model together with the 3D radiative transfer code SOC following the approach of \citet{schirmer2020}. We consider the density profile towards the Orion Bar described in \citet{Arab12} which agrees with the dust observational constraints from Spitzer and Herschel. A radiation field corresponding to a star of $T_{eff}$~=~40,000~K with $G_0\,=\,2,25\cdot10^4$ in Habing units is used. The  dust size distributions were adjusted in order to reproduce the IR observations of the Orion Bar (Spitzer IRAC 3.6, 4.5, 5.8 and 8~$\mu$m, Herschel PACS, 70 and 160~$\mu$m, and SPIRE 250, 350, and 500~$\mu$m):  the amorphous hydrocarbon nano-particle to gas ratio is set to about 100 times lower and their minimum size is set to 1.8 times larger than in the diffuse ISM \citep{Schirmer2022}. 
For the model template spectra of the atomic and H/H$_2$ transition region (Fig. \ref{fig:spectra}c), 
we calculate the average of the emission from the edge of the PDR up to a column density, from the edge, of 4.6 10$^{21}$ H cm$^{-2}$,
corresponding to the depth of the C/CO transition as computed by the PDR Meudon code (see above). For the molecular region (Fig. \ref{fig:spectra}d), we calculate the average of the emission from the layers which begins after the C/CO transition and ends at a density column from the edge of the PDR of 1.6 10$^{22}$ H cm$^{-2}$. For the calculation of the dust continuum and scattered light in the ionized region and ionization front, we used the DustEM optically thin model \citep{compiegne}. 
The incident spectrum on the \HII~region and the transmitted spectrum in the ionization front calculated by Cloudy was used. 
\\
{\bf Aromatic infrared band emission.} 
For each region, we also define a normalized (over the area) spectrum for AIB emission. These are computed 
using the four template spectra extracted by \citet{foschino2019} on ISO SWS data using machine learning
algorithms. The four spectra represent four families of PAH related species, namely  
neutral PAHs (PAH$^0$), cationic PAHs (PAH$^+$), evaporating very small grains (eVSGs) and 
large ionized PAHs (PAH$^x$). The description of these species is given in \citet{pilleri12, pilleri2015}
and \citet{foschino2019}. For the \HII{} region, the PAH spectrum consists completely of 
PAH$^0$, since the abundance of electrons is large and hence recombination of cationic 
PAHs is very efficient at keeping molecules neutral. 
At the ionization front, PAH emission 
is dominated by the neutral side of the interface, where density is highest, where gas is mostly neutral,
and where radiation field is high. Therefore, recombination with electrons is highly ineffective and only cationic PAHs, in particular the largest, can survive. This type of environment
is similar to that of planetary nebulae where PAH$^x$ are dominant \citep{joblin_carriers_2008}, hence we use this spectrum
for the ionization front. On the dissociation front, emission is dominated by neutral PAHs
but with a contribution of cations \citep[see e.g.][]{berne_blind_2012}, hence we adopt a spectrum consisting 
of 60$\%$ PAH$^{0}$ and 40$\%$ PAH$^+$. Finally, the molecular cloud region is dominated 
by eVSG and neutral PAHs, as seen in e.g. NGC 7023 \citep{berne_blind_2012}, and we adopt a spectrum with 
half eVSG and half PAH$^{0}$. 
The normalized AIB spectra were scaled to reproduce the integrated fluxes as observed in the Orion bar by Spitzer.
\\
{\bf Absolute calibration and recommended use of the spectra.} 
Absolute calibration of the total spectrum resulting from the sum of the four different regions has been cross-checked with existing observations of the Orion Bar (ISO, Spitzer, SOFI/NTT, HST), so as to confirm that the total spectrum is roughly realistic in terms of flux units. Moreover, for some specific gas lines (e.g., [FeII], H$_2$) observed recently with the Keck telescope  at very high angular resolution ($\sim$0.1$\arcsec$), 
we compare our model predictions to  the observed peak emission at the ionization front and dissociation front \citep{Habart2022}. This allowed us to ensure that we do not underestimate the peak emission by important factors. We emphasize that the individual spectra of the different regions, obtained by making simple assumptions and by separately estimating the different components (ionized gas, neutral gas, dust in ionized gas, dust in neutral gas, PAHs) 
are not fully realistic. 
However these spectra can be useful for time estimates with the JWST time exposure calculator, or testing data-analysis tools, before obtaining actual data.

\section{Observations} 
\label{sec:obs}

\begin{table}
    \caption{Filter selection for NIRCam and MIRI.}
    \label{tab:filters}
    \begin{center}
    \begin{tabular}{lccccc}
    \hline
    \hline
    Species & $\lambda~^1$ & Filter & Cont. & OB$^3$ & NGC\\
    & ($\mu$m) & & filter$^2$ & & 1982$^3$\\
    \hline
    \multicolumn{6}{c}{\textbf{NIRCam}}\\
     $\left[\mathrm{FeII}\right]~$   &  1.644 & F164N &  F162M & $\surd$ & \\
      Pa~$\alpha$  & 1.876 & F187N & F182M & $\surd$ & $\surd$\\
      Br~$\alpha$ & 4.052 & F405N & F410M & $\surd$ & $\surd$ \\
      H$_2$ & 2.120 & F212N & F210M & $\surd$ & $\surd$\\
      H$_2$ & 3.235 & F323N$^4$ & F300M & $\surd$ & \\
      H$_2$ & 4.694 & F470N & F480M & $\surd$ & \\
      AIB & 3.3 & F335M & F300M & $\surd$ & $\surd$\\
      & 1.405 & F140M & & $\surd$ & \\
      & 1.659 & F150W2 & & $\surd$ & \\
      & 2.672 & F277W & & $\surd$ & \\
      & 3.232 & F322W2 & & $\surd$ & \\
      & 4.408 & F444W & & $\surd$ & $\surd$\\[5pt]
    \hline
     \multicolumn{6}{c}{\textbf{MIRI}}\\
     & 7.7 & F770W & & $\surd$ & \\
     & 11.3& F1130W & & $\surd$ & \\
     & 15.0& F1500W & & $\surd$ & \\
     & 25.5& F2550W & & $\surd$ & \\[5pt]
    \hline
   \end{tabular}
    \end{center}
    $^1$ The wavelength of the transition or the pivot wavelength of the filter for NIRCam broad band filters. $^2$ Continuum filter; $^3$ OB: Orion Bar (on-target observations); NGC 1982: NIRCam parallel observations; $^4$ This filter is not contaminated by the 3.3 $\mu$m AIB.
\end{table}

The program objectives (Sect.~\ref{sec:immediate-goals}, ~\ref{sec:science-objectives}) require the use of near- and mid-IR spatially resolved spectroscopy to extract 
the spectra and signatures of the critical sub-regions in a PDR (Figs.~\ref{fig:pdr}, \ref{fig:overlay_spectrosopy}),
and near- and mid-IR imaging to obtain a general understanding of the environment. Hereafter we describe in details 
the set of planned observations as part of this ERS project with MIRI, NIRSpec and NIRCam. 
For the imaging, the filter selection is given in Table~\ref{tab:filters}. The file containing the 
details of these observations can be downloaded directly in the Astronomer Proposal Toolkit (APT)
using the program ID 1288.

\begin{figure}
\begin{center}
\includegraphics[scale=0.235]{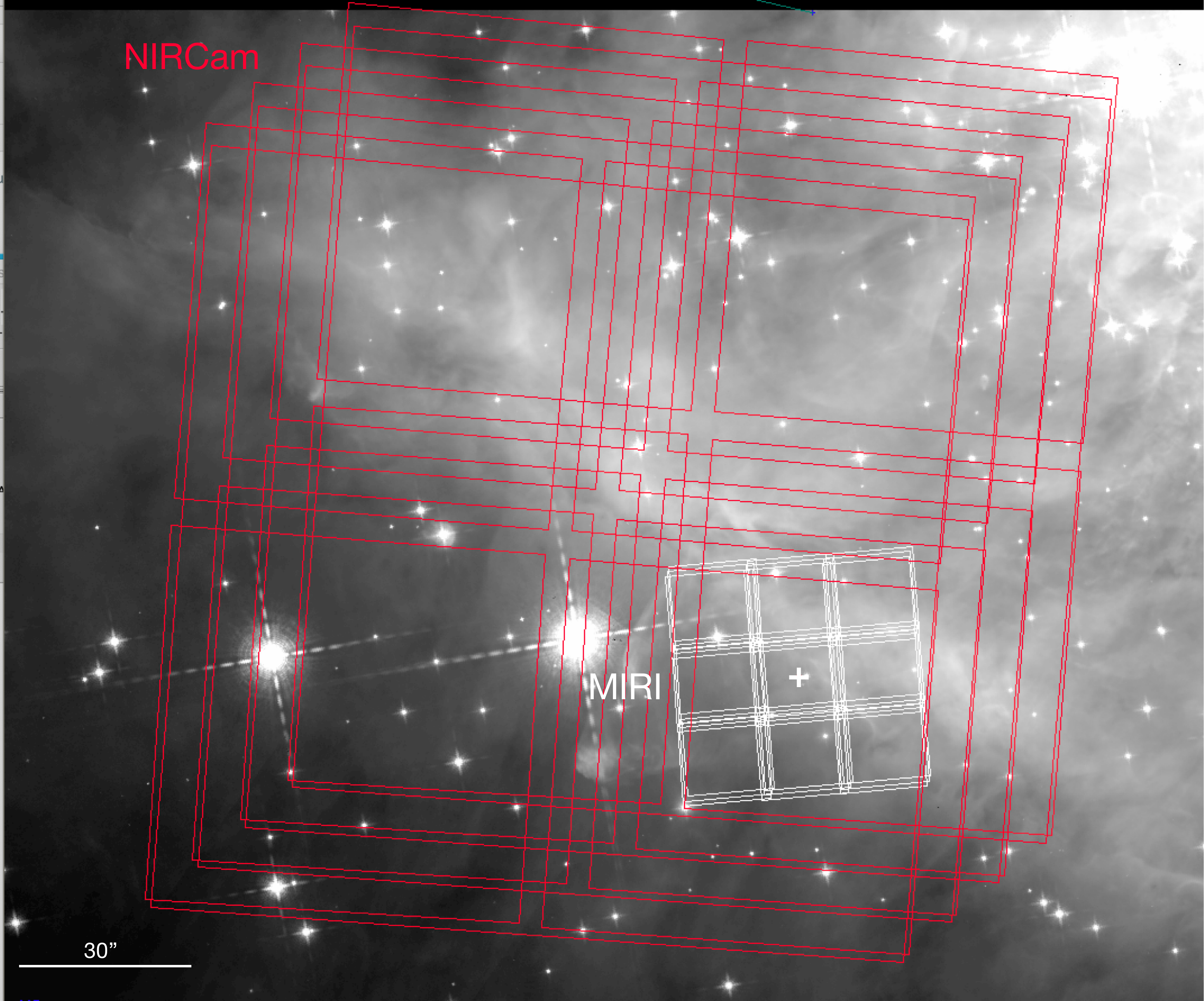}  
\end{center}
\vspace*{0cm}
\caption{Overlay of the JWST NIRCam (red) and MIRI (white) imaging footprints on the Hubble Space Telescope image of the Orion Bar at 1.3 $\mu$m \citep{rob20}. MIRI observations are centered on position $\alpha=$ 05 35 20.3448 and $\delta=$ -05 25 4.01 (position indicated by the white cross). For NIRCam, only module B covering the Bar is shown in this overlay. Module A is situated to the North (outside of image), and the pointing position (also outside this image) is situated in between these two modules at $\alpha=$ 05 35 20.1963, $\delta=$ -05 23 10.45. }
\label{fig:overlay_imaging}
\end{figure}

\subsection{NIRSpec IFU spectroscopy} 

We will obtain a $9\times1$ mosaic centered on position $\alpha=$ 05 35 20.4864,  $\delta=$ -05 25 10.96
(Fig.~\ref{fig:overlay_spectrosopy}) for the 1-5.3 $\mu$m range. The exact position angle (PA) for the mosaic will 
depend on the date of observation, however we plan to have PA $\sim 60^{\circ}$.
We use the H grisms with a spectral resolution of $\sim$2700, the NRSRAPID readout mode, which is appropriate 
for bright sources, and a 4 point dither. We include a dark exposure to quantify the leakage of the Micro-Shutter 
Array (MSA)\footnote{\url{https://jwst-docs.stsci.edu/jwst-near-infrared-spectrograph/nirspec-operations/nirspec-ifu-operations/nirspec-msa-leakage-correction-for-ifu-observations}}. We use five groups per integration with one integration and one exposure for a total on-source 
integration time of 257.7s\footnote{Definitions of groups, integrations, and exposures can be found at \url{https://jwst-docs.stsci.edu/understanding-exposure-times}.}. 
The expected SNRs have been computed using the Exposure Time Calculator (ETC)\footnote{\url{https://jwst.etc.stsci.edu/}}
provided by STScI, based on a reference spectrum which is the average of the four template spectra described in
Sect.~\ref{sec:simulated-spectra}. This provides a SNR per spectral resolution element on the continuum of at least 25, and up to several hundred 
for the 3.3 $\mu$m PAH band and up to a few thousand on bright lines.  In \citet{canin2021b} we provide a detailed 
simulation of the NIRSpec-IFU observations of the Orion Bar. 

\subsection{MIRI IFU spectroscopy}

We will obtain a $7\times2$ mosaic centered on position $\alpha=$ 05 35 20.4869, $\delta=$ -05 25 11.02
(Fig.~\ref{fig:overlay_spectrosopy}) over the entire 5-28.5 $\mu$m range, also with a PA $\sim 60^{\circ}$. 
We note that the NIRspec IFU spectroscopy mosaic described above is fully included in the MIRI IFU mosaic, 
the latter being larger, especially at longer wavelengths due to instrument design \citep{rieke15}.
We use the MRS spectrometer with a spectral resolution of $\sim$3000. We apply a 4 point dither 
optimized for extended sources and use the FASTR1 readout mode and the SUB128 imager subarray, both 
adapted for bright sources. We integrate 50.5s on-source using five groups per integration with 
one integration and one exposure. According to the ETC, using our reference spectrum this yields 
a minimal SNR per spectral resolution element of $\sim 10$ on the continuum, and up to above 500 for bright lines and PAH bands.

\subsection{NIRCam imaging} 

We will observe i) the 3.3 $\mu$m PAH band which, when combined with the 11.3 $\mu$m PAH band, measures PAH size \citep[Fig.~\ref{fig:diagnostics}c;][]{Ricca12, croiset, Knight21}, ii) the vibrationally excited lines of H$_2$ at 2.12, 3.23, and 4.7 $\mu$m, tracing the dissociation front, iii) the [FeII] at 1.64 $\mu$m, tracing the ionization front, and iv) the Pa~$\alpha$ and Br~$\alpha$ lines, tracing the \HII\, region. For each, we include a reference filter for the subtraction of the underlying continuum emission. We will also obtain broad band observations at 1.5, 2.7, 3.2, and 4.4 $\mu$m. We will map the PDR region with a single pointing using a 4 point dither (Fig.~\ref{fig:overlay_imaging}). All used filters are summarized in Table~\ref{tab:filters}.
We use the RAPID readout mode since the Orion Bar is very bright. We will obtain a SNR on the extended emission above 10 for all filters with a total on-source exposure time of 85.9s (2 groups per integration, one integration and one exposure).

\subsection{MIRI imaging} 

We will observe i) the 7.7 $\mu$m PAH band, ii) the 11.3 $\mu$m PAH band which, when combined,
provide a proxy for PAH ionization \citep[e.g.,][]{Joblin:96, hony_ch_2001}, iii) VSGs \citep[][]{desert}, and iv) continuum emission
at 25 $\mu$m, tracing warm dust in the \HII\, region, similarly to the corresponding WISE, Spitzer, and IRAS filters.
We obtain a $3\times3$ mosaic using 2 dithers and a 2 point dither pattern (2-POINT-MIRI-F770W-WITH-NIRCam) with 4
pointings  (Fig. \ref{fig:overlay_imaging}). To accommodate the brightness of the Orion Bar, we use the FASTR1 reading
mode and the SUB128 imager subarray. 
We obtain a SNR above 10 with a total on-source integration time of 136.9s (5 groups per integration, 115
integrations, and 1 exposure). 

\subsection{NIRCam parallel observations} 

We will obtain parallel NIRCam observations with the on-source MIRI imaging. 
The exact FOV will depend on the time of observations. 
If the observations are taken in the September visibility
window (as we expect as of today), it will be located North of the Orion Bar, on the NGC 1982 region. 
We will observe i) the 3.3 $\mu$m PAH band, 
ii) the vibrationally excited line of H$_2$ at 2.12 $\mu$m, and iii) the Pa~$\alpha$ and Br~$\alpha$ lines. 
For each, we include a reference filter for the subtraction of the underlying continuum emission. 
We will also obtain a broad band observation at 4.4 $\mu$m. The filters for theses
observations are summarized in Table~\ref{tab:filters}.
The pointings, number of dithers, 
and dither pattern are set by those of the primary observations (on-source MIRI imaging). 
To accommodate the brightness of the Orion Bar, we will use the BRIGHT2 readout mode and will obtain 
a total integration time of 128.8s (using 2 groups per integration, 1 integration, 
and 1 exposure).

\subsection{``Off" observations} 

We include spectroscopic ``off" observations to subtract any undesirable 
emission from scattered light from the Sun or the Galactic plane, telescope emission, Zodiacal light, 
or instrumental signal. This ensures the highest quality data, and provides information on possible 
background sources. 
We have chosen a background position (same for both instruments) situated in a region of very 
low emission in 2MASS and WISE surveys at a distance of 2 degrees from our on-source observations at $\alpha =$ 05 27 19.40 and $\delta =$ -05 32 4.40.

\subsection{Risks of saturation: estimates and mitigation}

We have conducted a number of tests to estimate and mitigate the risks of saturation. 

For the spectroscopic observations, the standard ETC calculations do not show any warning of saturation when employing a reference spectrum of the Orion Bar. However, the true (peak) intensities could be underestimated due to the increased spatial and spectral resolution of JWST compared to prior observations. We therefore took a conservative approach and investigated potential risks when employing this reference spectrum multiplied by a factor of 3. For MIRI, we obtain partial saturation\footnote{Partial saturation indicates saturation occurs in the third group or a later group of the integration ramp (\url{https://jwst-docs.stsci.edu/jwst-exposure-time-calculator-overview/jwst-etc-calculations-page-overview/jwst-etc-saturation}).} in the [\NeII{}] line at 12.8 $\mu$m and in the [\SIII{}] line at 18.7~$\mu$m. This risk is thus mainly affecting the part of the mosaic situated in the \HII{} region and only concerns theses lines (i.e. the rest of the spectrum is not saturated in this exercise). Since the saturation is only partial and the integration ramps are usable up to when the saturation occurs, we expect that fluxes can still be recovered for these lines. We do not see any saturation issues with NIRSpec. 
Regarding the photometric observations, we apply the same method to probe potential risks. For NIRCam observations, the ETC calculations do not show any warnings using the reference spectrum nor this spectrum multiplied by a factor of 3 (assuming an extended source). Discussions with instrument teams suggest that saturation with NIRCam on the Orion Bar in the extended emission is highly unlikely. We have also run detailed modeling of the observations using the MIRAGE instrument simulator \footnote{\url{https://github.com/spacetelescope/mirage}}, inserting background images of the Orion Bar at appropriate wavelengths from Hubble or {\it Spitzer} \citep{canin2021}. These simulated data do not show any saturation in the extended emission. For saturation of point sources, we ran the MIRAGE simulator inserting a point source with the magnitude of $\theta^2$ Ori A \citep{canin2021}. The star saturates the detector but only in a few pixels at the position of the star. Stray light is also limited to the immediate surroundings of the star. Overall, our predictions suggest that the risk of saturation for NIRCam on the Orion Bar is low. For MIRI imaging, the ETC predicts that the Orion Bar saturates the detectors in all filters if the full array is used for an extended source with the reference spectrum multiplied by 3. We have therefore modified the observations
and adopted a strategy which consists in reading the detector much faster only on a sub-field of 128$\times$128 pixels (SUB128) to avoid saturation. The resulting field of view is thus much smaller, hence to cover a large enough region, we must perform a $3\times3$ mosaic (Fig.~\ref{fig:overlay_imaging}). This is more time consuming that a standard strategy, however this is the only approach to limit the risks of saturation.

\section{Science-Enabling Products (SEPs)}
\label{sec:SEPs}

In the context of this ERS project, we will provide three types of SEPs to the community: 1) enhanced data products, 2) products facilitating data reduction and processing, and 3) data-interpretation tools via STScI and our website\footnote{\url{http://pdrs4all.org}}. First, we will provide the following enhanced data products:
\begin{itemize}
    \item \textbf{Highly processed data products} of the obtained observations. We will provide the data in post-pipeline format (Stage 3). We will also provide highly processed data, in particular   stitched IFU cubes for NIRSpec and MIRI.  
    \item \textbf{Maps of spectral features}. Using the spectral decomposition tool PAHFIT \citep[see below,][]{smith}, we will produce integrated line and band maps, with uncertainties, from our IFU spectroscopy observations.  Final products will be maps (FITS files) in physical units and astrometry.
    \item \textbf{Template spectra from the key regions} (the \HII\, region, the ionization front, the dissociation front, and the molecular zone). These will be extracted directly from the observations and by blind signal separation methods \citep[e.g.][]{berne_blind_2012}. The templates can be used by the community to interpret spectroscopic observations of non-resolved PDRs, such as in distant galaxies or proto-planetary disks.  
\end{itemize}

Second, to facilitate post-pipeline data reduction and processing of similar JWST observations, we will deliver:
\begin{itemize}
\item  \textbf{Spectral order stitching and stitched cubes.}
The official JWST pipeline assumes that combining spectra taken with different gratings and/or settings can be done without any specific matching step. This is a success-oriented view, however past experience (with e.g. Spitzer-IRS, ISO-SWS) has shown that extra corrections are usually needed to provide smooth transitions between such spectral segments. We will quantify any spectral stitching issues. If needed, we will develop a Python-based tool to provide the needed extra corrections and, if possible, make it automated. 
\item \textbf{Cross-correlation of spectra \& images.}
For the overlapping photometric and spectroscopic FOVs, a comparison will be done between the flux in specific (spectroscopic) lines or bands and the flux observed in specific (photometric) filters or combinations of filters (narrow and/or wide). This will be used for cross calibration purposes and to define color correction for fields with bright PAH bands \citep[e.g.][]{reach}. The products will be correlation tables and simple mathematical recipes. 

\item \textbf{PAHFIT.} This tool decomposes a spectrum into gas lines, dust features (aromatic/PAHs, aliphatics, fullerenes, silicates, ices), and dust continuum components and was developed for the analysis of Spitzer-IRS observations \citep{smith}. Recently, \citet{Lai:20} applied it to AKARI-Spitzer observations. We will extend its capabilities to include JWST applicability, various decomposition methods for dust features and continuum, and various extinction determination methods (silicate absorption, H$_2$ lines) so it becomes more versatile. In addition, we will make sure it can automatically treat all pixels in IFU maps.  This is implemented as a community open-source Python project
 (\url{https://github.com/PAHFIT/pahfit}).
 
\item \textbf{Line list} of all the lines and bands present in the data. The final product will be an ASCII table with the wavelength position, characteristic (line or band), and assignment of each line. For the lines, upper energy levels and Einstein coefficients for spontaneous emission will be added.
\end{itemize}

We will also provide an extensive and powerful set of tools for the interpretation of JWST data:
\begin{itemize}
\item \textbf{H$_{2}$ fitting tool} will provide fits of observed ${\rm H_2}$ excitation diagrams (e.g., Fig.~\ref{fig:diagnostics}) for any pixel and yield the warm and hot (UV-pumped) excitation temperatures (${\rm T_{ex}}$), warm, hot, and total column densities (${\rm N_{H_2}}$), and ortho-to-para ratio (${\rm R_{otp}}$). Outputs will be excitation diagrams and maps of ${\rm T_{ex}}$, ${\rm N_{H_2}}$ and ${\rm R_{otp}}$. 
The tool will be part of the PDR Toolbox \cite[\url{http://dustem.astro.umd.edu}]{2008ASPC..394..654P}, which has been rewritten as an open-source Python package (\url{https://pdrtpy.readthedocs.io}).

\item \textbf{The interstellar medium database, ISMDB,} is a model database with a web-based fitting tool to search in massive grids of state-of-the-art PDR models and derive physical parameters from observations of any number of spectral lines to be observed with JWST (\url{http://ismdb.obspm.fr}). This web service will give access to model grids from the Meudon PDR code \citep{lepetit} and the KOSMA-tau PDR code \citep{markus}. This tool will be able to handle maps of multiple lines (rather than single pointing observations in the existing version) and return the estimated parameter maps, providing in addition maps of the goodness-of-fit at each pixel and of the uncertainties at each pixel, as well single-pixel analysis tools allowing to visualize uncertainty correlations between the different parameters.

\item \textbf{pyPAHdb Spectral Analysis Tool}  decomposes a PAH emission spectrum into contributing sub-populations \citep[e.g., charge, size, composition, structure;][]{Shannon:18}. It is based on tools provided via the NASA Ames PAH IR Spectroscopic Database version 3.2 \citep[PAHdb; \url{http://www.astrochemistry.org/pahdb/}, ][]{Cami:11, bauschlicher, boersma:db, bauschlicher2, Mattioda:20}, i.e., the AmesPAHdbIDLSuite and the AmesPAHdbPythonSuite (\url{https://github.com/PAHdb/}). The tool makes use of a matrix of pre-calculated emission spectra from version 3.20 of PAHdb's library of density-functional-theory computed absorption cross-sections. The products will be maps of the PAH ionization and large fraction. The tool is implemented as a community open-source Python project
 (\url{https://github.com/PAHdb/pyPAHdb}).

\item \textbf{Ionized gas lines diagnostic diagrams.} 
We will provide diagnostic diagrams of key species to be observed with JWST for the interpretation of ionized gas lines (Fig.~\ref{fig:diagnostics}) and its conversion into physical conditions and extinction. These diagrams will be based on multi-level models or Cloudy \citep{cloudy} and can be used for sources showing emission of ionized gas. This tool has recently been implemented in the PDR Toolbox (\url{http://dustem.astro.umd.edu}).

\end{itemize}

\section{Community}
\label{sec:team}
\begin{figure*}[!th]
    \centering
    \includegraphics[clip,trim =0cm 8cm .5cm 5cm, scale=0.65]{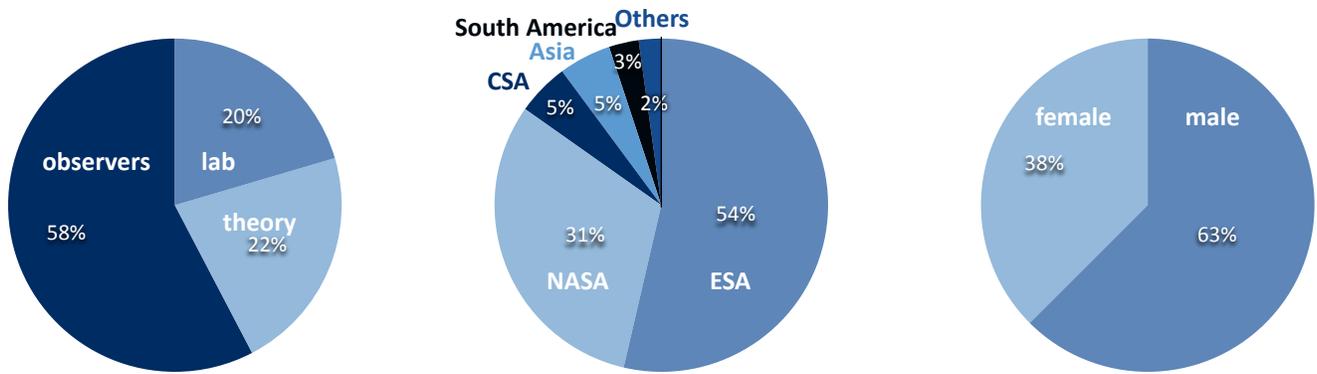}
    \vspace*{-.25cm}
    \caption{Detailed demographics of the PDRs4All team. The right chart represents the gender balance based on first names as given by the python package gender-guesser0.4.0. For the first names found in this database (90\% of the team),  results `female' and `mostly\_female' were combined so were the results `male' and `mostly\_male' while the result `androgynous' (1 name) was equally split over both genders.}
    \label{fig:demographics}    
\end{figure*}

The philosophy of this ERS program has been, from the start, to be open to the largest possible number of scientists, with the objective to gather together the international community. This has materialized into a large international, interdisciplinary, and diverse team of 145 scientists from 18 countries. The team includes observers, theoreticians, and experimentalists in the fields of astronomy, physics, and
chemistry, and has a gender balance of 41\% women (Fig.~\ref{fig:demographics}). The program is led by the PI team (O. Bern\'{e}, E. Habart \& E. Peeters) who are assisted by a core team of 17 scientists\footnote{The core team consists of A. Abergel, E. Bergin, J. Bernard-Salas, E. Bron, J. Cami, S. Cazaux, E. Dartois, A. Fuente, J. R. Goicoechea, K. Gordon,  Y. Okada, T. Onaka, M. Robberto, M. R\"ollig, A. Tielens, S. Vicente, M. Wolfire.}. This core team is complemented by extended-core team members who significantly contribute to the program, in particular with respect to the Science-Enabling Products.  
Telecons open to the community will be organized on a regular basis to disseminate data reduction and analysis techniques and recipes, best practices to design JWST proposals, and tutorials on the provided SEPs. We will organize the workshop ``Galactic and extragalactic PDRs with JWST". For more information, please visit \url{http://pdrs4all.org}.

\section{Conclusions}
\label{sec:conclusions}

This paper presents the JWST ERS program PDRs4All: ``Radiative feedback from massive stars". PDRs4All will be {\it the} go-to place for JWST observers who 1) are eager to see one of the first data taken with JWST, 2) are interested in how JWST's instruments perform on bright, extended targets, 3) need advanced software tools for the processing, analysis, and interpretation of JWST data, or 4) are simply interested in PDR physics and chemistry. Indeed, the unprecedented capabilities of JWST's instruments will give access to the key physical and chemical processes present in PDRs on the scale at which these processes occur. Therefore, without any exaggeration, PDRs4All will revolutionize our understanding of PDRs and provide major insight for the interpretation of unresolved PDRs, such as the surfaces of proto-planetary disks or distant star-forming galaxies. 

\section{Acknowledgement}
We are grateful to the PAHFIT developers team (Karl Gordon, Thomas Lai, Alexandros Maragkoudakis, Els Peeters, Bethany Schefter, Ameek Sidhu, and J.D.\ Smith).

Support for JWST-ERS program ID 1288 was provided through grants from the STScI under NASA contract NAS5-03127 to STScI (K.G., D. VDP., M.R.), Univ.\ of Maryland (M.W., M.P.), Univ.\ of Michigan (E.B., F.A.), and Univ.\ of Toledo (T.S.-Y.L.).

O.B. and E.H. are supported by the Programme National “Physique et Chimie du Milieu Interstellaire” (PCMI) of CNRS/INSU with INC/INP co-funded by CEA and CNES, and through APR grants 6315 and 6410 provided by CNES. 
E.P. and J.C. acknowledge support from the National Science and Engineering Council of Canada (NSERC) Discovery Grant program (RGPIN-2020-06434 and RGPIN-2021-04197 respectively). E.P. acknowledges support from a Western Strategic Support Accelerator Grant (ROLA ID 0000050636).
J.R.G. and S. C. thank the Spanish MCINN for funding support under grant \mbox{PID2019-106110GB-I00}.
Work by M.R. and Y.O. is carried out within the Collaborative Research Centre 956, sub-project C1, funded by the Deutsche Forschungsgemeinschaft (DFG) – project ID 184018867.
T.O. acknowledges support from JSPS Bilateral Program, Grant Number 120219939.
M.P. and M.W. acknowledge support 
from NASA Astrophysics Data Analysis Program award \#80NSSC19K0573. 
C.B. is grateful for an appointment at NASA Ames Research Center through the San Jos\'e State University Research Foundation (NNX17AJ88A) and acknowledges support from the Internal Scientist Funding Model (ISFM) Directed Work Package at NASA Ames titled: ``Laboratory Astrophysics -- The NASA Ames PAH IR Spectroscopic Database''.  


\bibliographystyle{aasjournal}
\bibliography{mainbib}

\begin{thebibliography}{}
\expandafter\ifx\csname natexlab\endcsname\relax\def\natexlab#1{#1}\fi
\providecommand{\url}[1]{\href{#1}{#1}}

\bibitem[{{Abel} {et~al.}(2006){Abel}, {Ferland}, {O'Dell}, {Shaw}, \&
  {Troland}}]{Abel06}
{Abel}, N.~P., {Ferland}, G.~J., {O'Dell}, C.~R., {Shaw}, G., \& {Troland},
  T.~H. 2006, \apj, 644, 344

\bibitem[{{Abgrall} {et~al.}(1992){Abgrall}, {Le Bourlot}, {Pineau Des Forets},
  {Roueff}, {Flower}, \& {Heck}}]{abgrall}
{Abgrall}, H., {Le Bourlot}, J., {Pineau Des Forets}, G., {et~al.} 1992, \aap,
  253, 525

\bibitem[{{Ag{\'u}ndez} {et~al.}(2010){Ag{\'u}ndez}, {Goicoechea},
  {Cernicharo}, {Faure}, \& {Roueff}}]{Agundez10}
{Ag{\'u}ndez}, M., {Goicoechea}, J.~R., {Cernicharo}, J., {Faure}, A., \&
  {Roueff}, E. 2010, \apj, 713, 662

\bibitem[{{Alata} {et~al.}(2014){Alata}, {Cruz-Diaz}, {Mu{\~n}oz Caro}, \&
  {Dartois}}]{Alata2014}
{Alata}, I., {Cruz-Diaz}, G.~A., {Mu{\~n}oz Caro}, G.~M., \& {Dartois}, E.
  2014, \aap, 569, A119

\bibitem[{{Allamandola} {et~al.}(1985){Allamandola}, {Tielens}, \&
  {Barker}}]{allamandola_polycyclic_1985}
{Allamandola}, L.~J., {Tielens}, A.~G.~G.~M., \& {Barker}, J.~R. 1985, \apjl,
  290, L25

\bibitem[{{Allers} {et~al.}(2005){Allers}, {Jaffe}, {Lacy}, {Draine}, \&
  {Richter}}]{Allers05}
{Allers}, K.~N., {Jaffe}, D.~T., {Lacy}, J.~H., {Draine}, B.~T., \& {Richter},
  M.~J. 2005, \apj, 630, 368

\bibitem[{{Anderson} {et~al.}(2012){Anderson}, {Zavagno}, {Deharveng},
  {Abergel}, {Motte}, {Andr{\'e}}, {Bernard}, {Bontemps}, {Hennemann}, {Hill},
  {Rod{\'o}n}, {Roussel}, \& {Russeil}}]{Anderson:12}
{Anderson}, L.~D., {Zavagno}, A., {Deharveng}, L., {et~al.} 2012, \aap, 542,
  A10

\bibitem[{{Andree-Labsch} {et~al.}(2017){Andree-Labsch}, {Ossenkopf-Okada}, \&
  {R{\"o}llig}}]{Andree17}
{Andree-Labsch}, S., {Ossenkopf-Okada}, V., \& {R{\"o}llig}, M. 2017, \aap,
  598, A2

\bibitem[{{Andrews} {et~al.}(2015){Andrews}, {Boersma}, {Werner}, {Livingston},
  {Allamandola}, \& {Tielens}}]{andrews}
{Andrews}, H., {Boersma}, C., {Werner}, M.~W., {et~al.} 2015, ApJ, 807, 99

\bibitem[{{Arab} {et~al.}(2012){Arab}, {Abergel}, {Habart}, {Bernard-Salas},
  {Ayasso}, {Dassas}, {Martin}, \& {White}}]{Arab12}
{Arab}, H., {Abergel}, A., {Habart}, E., {et~al.} 2012, \aap, 541, A19

\bibitem[{{Bakes} \& {Tielens}(1994)}]{bakesandtielens94}
{Bakes}, E.~L.~O., \& {Tielens}, A.~G.~G.~M. 1994, \apj, 427, 822

\bibitem[{{Balick} {et~al.}(1974){Balick}, {Gammon}, \& {Hjellming}}]{Balick74}
{Balick}, B., {Gammon}, R.~H., \& {Hjellming}, R.~M. 1974, \pasp, 86, 616

\bibitem[{{Bally} {et~al.}(2000){Bally}, {O'Dell}, \&
  {McCaughrean}}]{bally2000}
{Bally}, J., {O'Dell}, C.~R., \& {McCaughrean}, M.~J. 2000, \aj, 119, 2919

\bibitem[{{Bauschlicher} {et~al.}(2018){Bauschlicher}, {Ricca}, {Boersma}, \&
  {Allamandola}}]{bauschlicher2}
{Bauschlicher}, Jr., C.~W., {Ricca}, A., {Boersma}, C., \& {Allamandola}, L.~J.
  2018, \apjs, 234, 32

\bibitem[{{Bauschlicher} {et~al.}(2010){Bauschlicher}, {Boersma}, {Ricca},
  {Mattioda}, {Cami}, {Peeters}, {S{\'a}nchez de Armas}, {Puerta Saborido},
  {Hudgins}, \& {Allamandola}}]{bauschlicher}
{Bauschlicher}, Jr., C.~W., {Boersma}, C., {Ricca}, A., {et~al.} 2010, ApJS,
  189, 341

\bibitem[{Bayet {et~al.}(2009)Bayet, Viti, Williams, Rawlings, \&
  Bell}]{bayet2009molecular}
Bayet, E., Viti, S., Williams, D., Rawlings, J., \& Bell, T. 2009, The
  Astrophysical Journal, 696, 1466

\bibitem[{Bell {et~al.}(2006)Bell, Roueff, Viti, \&
  Williams}]{bell2006molecular}
Bell, T., Roueff, E., Viti, S., \& Williams, D. 2006, Monthly Notices of the
  Royal Astronomical Society, 371, 1865

\bibitem[{{Bernard-Salas} {et~al.}(2009){Bernard-Salas}, {Peeters}, {Sloan},
  {Gutenkunst}, {Matsuura}, {Tielens}, {Zijlstra}, \&
  {Houck}}]{Bernard-Salas:09}
{Bernard-Salas}, J., {Peeters}, E., {Sloan}, G.~C., {et~al.} 2009, \apj, 699,
  1541

\bibitem[{{Bernard-Salas} \& {Tielens}(2005)}]{bernard2005physical}
{Bernard-Salas}, J., \& {Tielens}, A.~G.~G.~M. 2005, \aap, 431, 523

\bibitem[{{Bernard-Salas} {et~al.}(2012){Bernard-Salas}, {Habart}, {Arab},
  {Abergel}, {Dartois}, {Martin}, {Bontemps}, {Joblin}, {White}, {Bernard}, \&
  {Naylor}}]{Bernard-Salas12}
{Bernard-Salas}, J., {Habart}, E., {Arab}, H., {et~al.} 2012, \aap, 538, A37

\bibitem[{{Bern{\'e}} {et~al.}(2017){Bern{\'e}}, {Cox}, {Mulas}, \&
  {Joblin}}]{berne_detection_2017}
{Bern{\'e}}, O., {Cox}, N.~L.~J., {Mulas}, G., \& {Joblin}, C. 2017, \aap, 605,
  L1

\bibitem[{Bern{\'e} {et~al.}(2012)Bern{\'e}, Joblin, Deville, Pilleri, Pety,
  Teyssier, Gerin, \& Fuente}]{berne_blind_2012}
Bern{\'e}, O., Joblin, C., Deville, Y., {et~al.} 2012, Proceedings of the
  annual meeting of the french society of astronomy and astrophysics (SF2A), 6.
\newblock \url{https://arxiv.org/abs/1210.3453}

\bibitem[{Bern{\'e} {et~al.}(2009)Bern{\'e}, Joblin, Fuente, \&
  M{\'e}nard}]{berne_what_2009}
Bern{\'e}, O., Joblin, C., Fuente, A., \& M{\'e}nard, F. 2009, \aap, 495, 827.
\newblock \url{http://www.aanda.org/10.1051/0004-6361:200810559}

\bibitem[{{Bern{\'e}} {et~al.}(2015){Bern{\'e}}, {Montillaud}, \&
  {Joblin}}]{berne2015}
{Bern{\'e}}, O., {Montillaud}, J., \& {Joblin}, C. 2015, \aap, 577, A133

\bibitem[{{Boersma} {et~al.}(2015){Boersma}, {Bregman}, \&
  {Allamandola}}]{boersma:15}
{Boersma}, C., {Bregman}, J., \& {Allamandola}, L.~J. 2015, \apj, 806, 121

\bibitem[{{Boersma} {et~al.}(2012){Boersma}, {Rubin}, \& {Allamandola}}]{boe12}
{Boersma}, C., {Rubin}, R.~H., \& {Allamandola}, L.~J. 2012, \apj, 753, 168

\bibitem[{{Boersma} {et~al.}(2014){Boersma}, {Bauschlicher}, {Ricca},
  {Mattioda}, {Cami}, {Peeters}, {S{\'a}nchez de Armas}, {Puerta Saborido},
  {Hudgins}, \& {Allamandola}}]{boersma:db}
{Boersma}, C., {Bauschlicher}, C.~W., J., {Ricca}, A., {et~al.} 2014, \apjs,
  211, 8

\bibitem[{{Bregman} {et~al.}(1989){Bregman}, {Allamandola}, {Tielens},
  {Geballe}, \& {Witteborn}}]{Bregman89}
{Bregman}, J.~D., {Allamandola}, L.~J., {Tielens}, A.~G.~G.~M., {Geballe},
  T.~R., \& {Witteborn}, F.~C. 1989, \apj, 344, 791

\bibitem[{{Bron} {et~al.}(2018){Bron}, {Ag{\'u}ndez}, {Goicoechea}, \&
  {Cernicharo}}]{bron18}
{Bron}, E., {Ag{\'u}ndez}, M., {Goicoechea}, J.~R., \& {Cernicharo}, J. 2018,
  arXiv e-prints, arXiv:1801.01547

\bibitem[{{Bron} {et~al.}(2014){Bron}, {Le Bourlot}, \& {Le Petit}}]{bron14}
{Bron}, E., {Le Bourlot}, J., \& {Le Petit}, F. 2014, \aap, 569, A100

\bibitem[{{Bron} {et~al.}(2016){Bron}, {Le Petit}, \& {Le Bourlot}}]{bron16}
{Bron}, E., {Le Petit}, F., \& {Le Bourlot}, J. 2016, \aap, 588, A27

\bibitem[{{Burton} {et~al.}(1990){Burton}, {Hollenbach}, \&
  {Tielens}}]{Burton90}
{Burton}, M.~G., {Hollenbach}, D.~J., \& {Tielens}, A.~G.~G.~M. 1990, \apj,
  365, 620

\bibitem[{{Burton} {et~al.}(1998){Burton}, {Howe}, {Geballe}, \&
  {Brand}}]{Burton:98}
{Burton}, M.~G., {Howe}, J.~E., {Geballe}, T.~R., \& {Brand}, P.~W.~J.~L. 1998,
  \pasa, 15, 194

\bibitem[{{Calzetti}(2020)}]{Calzetti:20}
{Calzetti}, D. 2020, Nature Astronomy, 4, 437

\bibitem[{{Cami}(2011)}]{Cami:11}
{Cami}, J. 2011, in EAS Publications Series, Vol.~46, EAS Publications Series,
  ed. C.~{Joblin} \& A.~G.~G.~M. {Tielens}, 117--122

\bibitem[{{Candian} {et~al.}(2012){Candian}, {Kerr}, {Song}, {McCombie}, \&
  {Sarre}}]{Candian:12}
{Candian}, A., {Kerr}, T.~H., {Song}, I.-O., {McCombie}, J., \& {Sarre}, P.~J.
  2012, \mnras, 426, 389

\bibitem[{Canin {et~al.}(2021)Canin, Berné, \& team}]{canin2021}
Canin, A., Berné, O., \& team, T. P.~E. 2021, PDRs4all: Simulation and data
  reduction of JWST NIRCam imaging of an extended bright source, the Orion Bar,
  , , arXiv:2112.03106

\bibitem[{Canin {et~al.}(2022)Canin, Berné, \& team}]{canin2021b}
---. 2022, PDRs4all: NIRSpec simulation of integral field unit spectroscopy of
  the Orion Bar photodissociation region, , , arXiv:2201.01092

\bibitem[{{Cardelli} {et~al.}(1989){Cardelli}, {Clayton}, \&
  {Mathis}}]{Cardelli89}
{Cardelli}, J.~A., {Clayton}, G.~C., \& {Mathis}, J.~S. 1989, \apj, 345, 245

\bibitem[{{Carlsten} \& {Hartigan}(2018)}]{Carlsten2018}
{Carlsten}, S.~G., \& {Hartigan}, P.~M. 2018, \apj, 869, 77

\bibitem[{{Castellanos} {et~al.}(2014){Castellanos}, {Bern{\'e}}, {Sheffer},
  {Wolfire}, \& {Tielens}}]{castellanos_c_2014_fc}
{Castellanos}, P., {Bern{\'e}}, O., {Sheffer}, Y., {Wolfire}, M.~G., \&
  {Tielens}, A.~G.~G.~M. 2014, \apj, 794, 83

\bibitem[{{Cernicharo}(2004)}]{Cernicharo_2004}
{Cernicharo}, J. 2004, \apjl, 608, L41

\bibitem[{{Cesarsky} {et~al.}(1996){Cesarsky}, {Lequeux}, {Abergel}, {Perault},
  {Palazzi}, {Madden}, \& {Tran}}]{Cesarsky1996}
{Cesarsky}, D., {Lequeux}, J., {Abergel}, A., {et~al.} 1996, \aap, 315, L305

\bibitem[{{Cesarsky} {et~al.}(2000){Cesarsky}, {Lequeux}, {Ryter}, \&
  {G{\'e}rin}}]{Cesarsky2000}
{Cesarsky}, D., {Lequeux}, J., {Ryter}, C., \& {G{\'e}rin}, M. 2000, \aap, 354,
  L87

\bibitem[{{Champion} {et~al.}(2017){Champion}, {Bern{\'e}}, {Vicente}, {Kamp},
  {Le Petit}, {Gusdorf}, {Joblin}, \& {Goicoechea}}]{Champion17}
{Champion}, J., {Bern{\'e}}, O., {Vicente}, S., {et~al.} 2017, \aap, 604, A69

\bibitem[{Chevance {et~al.}(2016)Chevance, Madden, Lebouteiller, Godard,
  Cormier, Galliano, Hony, Indebetouw, Le~Bourlot, Lee,
  {et~al.}}]{chevance2016milestone}
Chevance, M., Madden, S., Lebouteiller, V., {et~al.} 2016, Astronomy \&
  Astrophysics, 590, A36

\bibitem[{{Chuss} {et~al.}(2019){Chuss}, {Andersson}, {Bally}, {Dotson},
  {Dowell}, {Guerra}, {Harper}, {Houde}, {Jones}, {Lazarian}, {Lopez
  Rodriguez}, {Michail}, {Morris}, {Novak}, {Siah}, {Staguhn}, {Vaillancourt},
  {Volpert}, {Werner}, {Wollack}, {Benford}, {Berthoud}, {Cox}, {Crutcher},
  {Dale}, {Fissel}, {Goldsmith}, {Hamilton}, {Hanany}, {Henning}, {Looney},
  {Moseley}, {Santos}, {Stephens}, {Tassis}, {Trinh}, {Van Camp},
  {Ward-Thompson}, \& {HAWC + Science Team}}]{Chuss19}
{Chuss}, D.~T., {Andersson}, B.~G., {Bally}, J., {et~al.} 2019, \apj, 872, 187

\bibitem[{{Compi{\`e}gne} {et~al.}(2008){Compi{\`e}gne}, {Abergel},
  {Verstraete}, \& {Habart}}]{compiegne2008}
{Compi{\`e}gne}, M., {Abergel}, A., {Verstraete}, L., \& {Habart}, E. 2008,
  \aap, 491, 797

\bibitem[{Compi{\`e}gne {et~al.}(2007)Compi{\`e}gne, Abergel, Verstraete,
  Reach, Habart, Smith, Boulanger, \& Joblin}]{compiegne_aromatic_2007}
Compi{\`e}gne, M., Abergel, A., Verstraete, L., {et~al.} 2007, \aap, 471, 205.
\newblock \url{http://www.aanda.org/10.1051/0004-6361:20066172}

\bibitem[{{Compi{\`e}gne} {et~al.}(2011){Compi{\`e}gne}, {Verstraete}, {Jones},
  {Bernard}, {Boulanger}, {Flagey}, {Le Bourlot}, {Paradis}, \&
  {Ysard}}]{compiegne}
{Compi{\`e}gne}, M., {Verstraete}, L., {Jones}, A., {et~al.} 2011, A\&A, 525,
  A103

\bibitem[{{Contreras} \& {Salama}(2013)}]{Contreras:13}
{Contreras}, C.~S., \& {Salama}, F. 2013, \apjs, 208, 6

\bibitem[{Cormier {et~al.}(2012)Cormier, Lebouteiller, Madden, Abel, Hony,
  Galliano, Baes, Barlow, Cooray, De~Looze, {et~al.}}]{cormier2012nature}
Cormier, D., Lebouteiller, V., Madden, S., {et~al.} 2012, Astronomy \&
  Astrophysics, 548, A20

\bibitem[{Cox {et~al.}(2015)Cox, Pilleri, Bern{\'e}, Cernicharo, \&
  Joblin}]{cox2015polycyclic}
Cox, N., Pilleri, P., Bern{\'e}, O., Cernicharo, J., \& Joblin, C. 2015,
  Monthly Notices of the Royal Astronomical Society: Letters, 456, L89

\bibitem[{{Croiset} {et~al.}(2016){Croiset}, {Candian}, {Bern{\'e}}, \&
  {Tielens}}]{croiset}
{Croiset}, B.~A., {Candian}, A., {Bern{\'e}}, O., \& {Tielens}, A.~G.~G.~M.
  2016, A\&A, 590, A26

\bibitem[{{Cuadrado} {et~al.}(2017){Cuadrado}, {Goicoechea}, {Cernicharo},
  {Fuente}, {Pety}, \& {Tercero}}]{Cuadrado17}
{Cuadrado}, S., {Goicoechea}, J.~R., {Cernicharo}, J., {et~al.} 2017, \aap,
  603, A124

\bibitem[{{Cuadrado} {et~al.}(2015){Cuadrado}, {Goicoechea}, {Pilleri},
  {Cernicharo}, {Fuente}, \& {Joblin}}]{Cuadrado15}
{Cuadrado}, S., {Goicoechea}, J.~R., {Pilleri}, P., {et~al.} 2015, \aap, 575,
  A82

\bibitem[{{Cuadrado} {et~al.}(2019){Cuadrado}, {Salas}, {Goicoechea},
  {Cernicharo}, {Tielens}, \& {B{\'a}ez-Rubio}}]{Cuadrado19}
{Cuadrado}, S., {Salas}, P., {Goicoechea}, J.~R., {et~al.} 2019, \aap, 625, L3

\bibitem[{{Cubick} {et~al.}(2008){Cubick}, {Stutzki}, {Ossenkopf}, {Kramer}, \&
  {R{\"o}llig}}]{cubick08}
{Cubick}, M., {Stutzki}, J., {Ossenkopf}, V., {Kramer}, C., \& {R{\"o}llig}, M.
  2008, \aap, 488, 623

\bibitem[{{Dartois} {et~al.}(2020){Dartois}, {Charon}, {Engrand}, {Pino}, \&
  {Sandt}}]{dartois20}
{Dartois}, E., {Charon}, E., {Engrand}, C., {Pino}, T., \& {Sandt}, C. 2020,
  Astronomy \& Astrophysics, 637, A82

\bibitem[{{Dere} {et~al.}(2019){Dere}, {Del Zanna}, {Young}, {Landi}, \&
  {Sutherland}}]{chianti}
{Dere}, K.~P., {Del Zanna}, G., {Young}, P.~R., {Landi}, E., \& {Sutherland},
  R.~S. 2019, \apjs, 241, 22

\bibitem[{D\'esert {et~al.}(1990)D\'esert, Boulanger, \&
  Puget}]{desert_interstellar_1990}
D\'esert, F.-X., Boulanger, F., \& Puget, J.-L. 1990, A{\textbackslash}\&A,
  237, 215

\bibitem[{{D\'esert} {et~al.}(1990){D\'esert}, {Boulanger}, \&
  {Puget}}]{desert}
{D\'esert}, F.-X., {Boulanger}, F., \& {Puget}, J.~L. 1990, A\&A, 237, 215

\bibitem[{{Doney} {et~al.}(2016){Doney}, {Candian}, {Mori}, {Onaka}, \&
  {Tielens}}]{doney2016}
{Doney}, K.~D., {Candian}, A., {Mori}, T., {Onaka}, T., \& {Tielens},
  A.~G.~G.~M. 2016, \aap, 586, A65

\bibitem[{Draine(2003)}]{draine_interstellar_2003}
Draine, B. 2003, Annual Review of Astronomy and Astrophysics, 41, 241.
\newblock \url{https://doi.org/10.1146/annurev.astro.41.011802.094840}

\bibitem[{{Draine}(1978)}]{draine1978}
{Draine}, B.~T. 1978, \apjs, 36, 595

\bibitem[{Draine \& Li(2007)}]{draine_infrared_2007}
Draine, B.~T., \& Li, A. 2007, The Astrophysical Journal, 657, 810.
\newblock \url{http://stacks.iop.org/0004-637X/657/i=2/a=810}

\bibitem[{{Esplugues} {et~al.}(2016){Esplugues}, {Cazaux}, {Meijerink},
  {Spaans}, \& {Caselli}}]{Esplugues16}
{Esplugues}, G.~B., {Cazaux}, S., {Meijerink}, R., {Spaans}, M., \& {Caselli},
  P. 2016, \aap, 591, A52

\bibitem[{{Fazio} {et~al.}(1974){Fazio}, {Kleinmann}, {Noyes}, {Wright},
  {Zeilik}, \& {Low}}]{Fazio74}
{Fazio}, G.~G., {Kleinmann}, D.~E., {Noyes}, R.~W., {et~al.} 1974, \apjl, 192,
  L23

\bibitem[{{Ferland} {et~al.}(2017){Ferland}, {Chatzikos}, {Guzm{\'a}n},
  {Lykins}, {van Hoof}, {Williams}, {Abel}, {Badnell}, {Keenan}, {Porter}, \&
  {Stancil}}]{cloudy}
{Ferland}, G.~J., {Chatzikos}, M., {Guzm{\'a}n}, F., {et~al.} 2017, \rmxaa, 53,
  385

\bibitem[{Fitzpatrick \& Massa(1990)}]{Fitzpatrick1990}
Fitzpatrick, E.~L., \& Massa, D. 1990, The Astrophysical Journal Supplement
  Series, 72, 163

\bibitem[{Foschino {et~al.}(2019)Foschino, Bern{\'e}, \& Joblin}]{foschino2019}
Foschino, S., Bern{\'e}, O., \& Joblin, C. 2019, Astronomy \& Astrophysics,
  632, A84

\bibitem[{{Fuente} {et~al.}(2003){Fuente}, {Rodr{\i}guez-Franco},
  {Garc{\i}a-Burillo}, {Mart{\i}n-Pintado}, \& {Black}}]{Fuente03}
{Fuente}, A., {Rodr{\i}guez-Franco}, A., {Garc{\i}a-Burillo}, S.,
  {Mart{\i}n-Pintado}, J., \& {Black}, J.~H. 2003, \aap, 406, 899

\bibitem[{{Galliano} {et~al.}(2008){Galliano}, {Madden}, {Tielens}, {Peeters},
  \& {Jones}}]{Galliano:08}
{Galliano}, F., {Madden}, S.~C., {Tielens}, A. G.~G.~M., {Peeters}, E., \&
  {Jones}, A.~P. 2008, \apj, 679, 310

\bibitem[{Gardner {et~al.}(2006)Gardner, Mather, Clampin, Doyon, Greenhouse,
  Hammel, Hutchings, Jakobsen, Lilly, Long, {et~al.}}]{gardner2006}
Gardner, J.~P., Mather, J.~C., Clampin, M., {et~al.} 2006, Space Science
  Reviews, 123, 485

\bibitem[{Gatchell {et~al.}(2021)Gatchell, Ameixa, Ji, Stockett, Simonsson,
  Denifl, Cederquist, Schmidt, \& Zettergren}]{Gatchell:21}
Gatchell, M., Ameixa, J., Ji, M., {et~al.} 2021, Nature Communications, 12,
  6646.
\newblock \url{https://doi.org/10.1038/s41467-021-26899-0}

\bibitem[{{Geballe} {et~al.}(1989){Geballe}, {Tielens}, {Allamandola},
  {Moorhouse}, \& {Brand}}]{geballe}
{Geballe}, T.~R., {Tielens}, A.~G.~G.~M., {Allamandola}, L.~J., {Moorhouse},
  A., \& {Brand}, P.~W.~J.~L. 1989, ApJ, 341, 278

\bibitem[{{Giard} {et~al.}(1994){Giard}, {Bernard}, {Lacombe}, {Normand}, \&
  {Rouan}}]{Giard94}
{Giard}, M., {Bernard}, J.~P., {Lacombe}, F., {Normand}, P., \& {Rouan}, D.
  1994, \aap, 291, 239

\bibitem[{{Goicoechea} \& {Cuadrado}(2021)}]{Goicoechea21}
{Goicoechea}, J.~R., \& {Cuadrado}, S. 2021, \aap, 647, L7

\bibitem[{Goicoechea {et~al.}(2015)Goicoechea, Teyssier, Etxaluze, Goldsmith,
  Ossenkopf, Gerin, Bergin, Black, Cernicharo, Cuadrado,
  {et~al.}}]{goicoechea2015}
Goicoechea, J.~R., Teyssier, D., Etxaluze, M., {et~al.} 2015, The Astrophysical
  Journal, 812, 75

\bibitem[{{Goicoechea} {et~al.}(2016){Goicoechea}, {Pety}, {Cuadrado},
  {Cernicharo}, {Chapillon}, {Fuente}, {Gerin}, {Joblin}, {Marcelino}, \&
  {Pilleri}}]{goicoechea}
{Goicoechea}, J.~R., {Pety}, J., {Cuadrado}, S., {et~al.} 2016, Nature, 537,
  207

\bibitem[{{Goicoechea} {et~al.}(2021){Goicoechea}, {Aguado}, {Cuadrado},
  {Roncero}, {Pety}, {Bron}, {Fuente}, {Riquelme}, {Chapillon}, {Herrera}, \&
  {Duran}}]{Goicoechea21b}
{Goicoechea}, J.~R., {Aguado}, A., {Cuadrado}, S., {et~al.} 2021, \aap, 647,
  A10

\bibitem[{Gorti {et~al.}(2009)Gorti, Dullemond, \& Hollenbach}]{gorti2009time}
Gorti, U., Dullemond, C., \& Hollenbach, D. 2009, The Astrophysical Journal,
  705, 1237

\bibitem[{{Gorti} \& {Hollenbach}(2002)}]{Gorti02}
{Gorti}, U., \& {Hollenbach}, D. 2002, \apj, 573, 215

\bibitem[{{Gould} \& {Salpeter}(1963)}]{gould1963}
{Gould}, R.~J., \& {Salpeter}, E.~E. 1963, \apj, 138, 393

\bibitem[{{Gro{\ss}schedl} {et~al.}(2018){Gro{\ss}schedl}, {Alves}, {Meingast},
  {Ackerl}, {Ascenso}, {Bouy}, {Burkert}, {Forbrich}, {F{\"u}rnkranz},
  {Goodman}, {Hacar}, {Herbst-Kiss}, {Lada}, {Larreina}, {Leschinski},
  {Lombardi}, {Moitinho}, {Mortimer}, \& {Zari}}]{Gross18}
{Gro{\ss}schedl}, J.~E., {Alves}, J., {Meingast}, S., {et~al.} 2018, \aap, 619,
  A106

\bibitem[{{G{\"u}del} {et~al.}(2008){G{\"u}del}, {Briggs}, {Montmerle},
  {Audard}, {Rebull}, \& {Skinner}}]{Gudel08}
{G{\"u}del}, M., {Briggs}, K.~R., {Montmerle}, T., {et~al.} 2008, Science, 319,
  309

\bibitem[{{Guerra} {et~al.}(2021){Guerra}, {Chuss}, {Dowell}, {Houde},
  {Michail}, {Siah}, \& {Wollack}}]{Guerra21}
{Guerra}, J.~A., {Chuss}, D.~T., {Dowell}, C.~D., {et~al.} 2021, \apj, 908, 98

\bibitem[{Guillard {et~al.}(2012)Guillard, Ogle, Emonts, Appleton, Morganti,
  Tadhunter, Oosterloo, Evans, \& Evans}]{guillard2012strong}
Guillard, P., Ogle, P., Emonts, B., {et~al.} 2012, The Astrophysical Journal,
  747, 95

\bibitem[{Guilloteau {et~al.}(2020{\natexlab{a}})Guilloteau, Oberlin,
  Bern{\'e}, \& Dobigeon}]{guilloteau2020b}
Guilloteau, C., Oberlin, T., Bern{\'e}, O., \& Dobigeon, N. 2020{\natexlab{a}},
  IEEE Transactions on Computational Imaging, 6, 1362

\bibitem[{Guilloteau {et~al.}(2020{\natexlab{b}})Guilloteau, Oberlin,
  Bern{\'e}, Habart, \& Dobigeon}]{guilloteau2020}
Guilloteau, C., Oberlin, T., Bern{\'e}, O., Habart, {\'E}., \& Dobigeon, N.
  2020{\natexlab{b}}, The Astronomical Journal, 160, 28

\bibitem[{{Guzm{\'a}n} {et~al.}(2011){Guzm{\'a}n}, {Pety}, {Goicoechea},
  {Gerin}, \& {Roueff}}]{Guzman11}
{Guzm{\'a}n}, V., {Pety}, J., {Goicoechea}, J.~R., {Gerin}, M., \& {Roueff}, E.
  2011, \aap, 534, A49

\bibitem[{{Guzm{\'a}n} {et~al.}(2015){Guzm{\'a}n}, {Pety}, {Goicoechea},
  {Gerin}, {Roueff}, {Gratier}, \& {{\"O}berg}}]{guzman2015}
{Guzm{\'a}n}, V.~V., {Pety}, J., {Goicoechea}, J.~R., {et~al.} 2015, \apjl,
  800, L33

\bibitem[{{Habart} {et~al.}(2011){Habart}, {Abergel}, {Boulanger}, {Joblin},
  {Verstraete}, {Compi{\`e}gne}, {Pineau Des For{\^e}ts}, \& {Le
  Bourlot}}]{Habart2011}
{Habart}, E., {Abergel}, A., {Boulanger}, F., {et~al.} 2011, \aap, 527, A122

\bibitem[{{Habart} {et~al.}(2004){Habart}, {Boulanger}, {Verstraete},
  {Walmsley}, \& {Pineau des For{\^e}ts}}]{habart04}
{Habart}, E., {Boulanger}, F., {Verstraete}, L., {Walmsley}, C.~M., \& {Pineau
  des For{\^e}ts}, G. 2004, \aap, 414, 531

\bibitem[{Habart {et~al.}(in prep)Habart, Le~Gal, Alvarez, Olivier~Berne,
  Peeters, Wolfire, Goicoechea, \& Wolfire}]{Habart2022}
Habart, E., Le~Gal, R., Alvarez, C., {et~al.} in prep

\bibitem[{{Habart} {et~al.}(2005){Habart}, {Walmsley}, {Verstraete}, {Cazaux},
  {Maiolino}, {Cox}, {Boulanger}, \& {Pineau des For{\^e}ts}}]{habart05}
{Habart}, E., {Walmsley}, M., {Verstraete}, L., {et~al.} 2005, \ssr, 119, 71

\bibitem[{{Habart} {et~al.}(2010){Habart}, {Dartois}, {Abergel}, {Baluteau},
  {Naylor}, {Polehampton}, {Joblin}, {Ade}, {Anderson}, {Andr{\'e}}, {Arab},
  {Bernard}, {Blagrave}, {Bontemps}, {Boulanger}, {Cohen}, {Compiegne}, {Cox},
  {Davis}, {Emery}, {Fulton}, {Gry}, {Huang}, {Jones}, {Kirk}, {Lagache},
  {Lim}, {Madden}, {Makiwa}, {Martin}, {Miville-Desch{\^e}nes}, {Molinari},
  {Moseley}, {Motte}, {Okumura}, {Pinheiro Gon{\c c}alves}, {Rodon}, {Russeil},
  {Saraceno}, {Sidher}, {Spencer}, {Swinyard}, {Ward-Thompson}, {White}, \&
  {Zavagno}}]{Habart10}
{Habart}, E., {Dartois}, E., {Abergel}, A., {et~al.} 2010, \aap, 518, L116

\bibitem[{{Habing}(1968)}]{Habing68}
{Habing}, H.~J. 1968, \bain, 19, 421

\bibitem[{{Hartigan} {et~al.}(2020){Hartigan}, {Downes}, \&
  {Isella}}]{Hartigan2020}
{Hartigan}, P., {Downes}, T., \& {Isella}, A. 2020, \apjl, 902, L1

\bibitem[{Helou {et~al.}(2001)Helou, Malhotra, Hollenbach, Dale, \&
  Contursi}]{hel01}
Helou, G., Malhotra, S., Hollenbach, D.~J., Dale, D.~A., \& Contursi, A. 2001,
  The Astrophysical Journal Letters, 548, L73

\bibitem[{{Herrmann} {et~al.}(1997){Herrmann}, {Madden}, {Nikola}, {Poglitsch},
  {Timmermann}, {Geis}, {Townes}, \& {Stacey}}]{Herrmann97}
{Herrmann}, F., {Madden}, S.~C., {Nikola}, T., {et~al.} 1997, \apj, 481, 343

\bibitem[{{Hogerheijde} {et~al.}(1995){Hogerheijde}, {Jansen}, \& {van
  Dishoeck}}]{Hoger95}
{Hogerheijde}, M.~R., {Jansen}, D.~J., \& {van Dishoeck}, E.~F. 1995, \aap,
  294, 792

\bibitem[{Hogerheijde {et~al.}(1995)Hogerheijde, Jansen, \& van
  Dishoeck}]{hogerheijde1995millimeter}
Hogerheijde, M.~R., Jansen, D.~J., \& van Dishoeck, E.~F. 1995, Astronomy and
  Astrophysics, 294, 792

\bibitem[{Hollenbach \& Natta(1995)}]{hollenbach1995time}
Hollenbach, D., \& Natta, A. 1995, The Astrophysical Journal, 455, 133

\bibitem[{{Hollenbach} \& {Tielens}(1999)}]{Hollenbach99}
{Hollenbach}, D.~J., \& {Tielens}, A.~G.~G.~M. 1999, Reviews of Modern Physics,
  71, 173

\bibitem[{Hony {et~al.}(2001)Hony, Van~Kerckhoven, Peeters, Tielens, Hudgins,
  \& Allamandola}]{hony_ch_2001}
Hony, S., Van~Kerckhoven, C., Peeters, E., {et~al.} 2001, \aap, 370, 1030.
\newblock \url{http://www.aanda.org/10.1051/0004-6361:20010242}

\bibitem[{Jansen {et~al.}(1995)Jansen, Spaans, Hogerheijde, \&
  Van~Dishoeck}]{jansen1995millimeter}
Jansen, D.~J., Spaans, M., Hogerheijde, M.~R., \& Van~Dishoeck, E.~F. 1995

\bibitem[{Joblin {et~al.}(2008)Joblin, Szczerba, Berné, \&
  Szyszka}]{joblin_carriers_2008}
Joblin, C., Szczerba, R., Berné, O., \& Szyszka, C. 2008, \aap, 490, 189.
\newblock \url{http://www.aanda.org/10.1051/0004-6361:20079061}

\bibitem[{Joblin \& Tielens(2011)}]{joblin2011pahs}
Joblin, C., \& Tielens, A. G. G.~M. 2011, PAHs and the Universe (EDP sciences)

\bibitem[{{Joblin} {et~al.}(1996{\natexlab{a}}){Joblin}, {Tielens},
  {Allamandola}, \& {Geballe}}]{Joblin:3umvsmethyl:96}
{Joblin}, C., {Tielens}, A.~G.~G.~M., {Allamandola}, L.~J., \& {Geballe}, T.~R.
  1996{\natexlab{a}}, \apj, 458, 610

\bibitem[{{Joblin} {et~al.}(1996{\natexlab{b}}){Joblin}, {Tielens}, {Geballe},
  \& {Wooden}}]{Joblin:96}
{Joblin}, C., {Tielens}, A.~G.~G.~M., {Geballe}, T.~R., \& {Wooden}, D.~H.
  1996{\natexlab{b}}, \apjl, 460, L119

\bibitem[{{Joblin} {et~al.}(2018){Joblin}, {Bron}, {Pinto}, {Pilleri}, {Le
  Petit}, {Gerin}, {Le Bourlot}, {Fuente}, {Bern{\'e}}, {Goicoechea}, {Habart},
  {K{\"o}hler}, {Teyssier}, {Nagy}, {Montillaud}, {Vastel}, {Cernicharo},
  {R{\"o}llig}, {Ossenkopf-Okada}, \& {Bergin}}]{Joblin18}
{Joblin}, C., {Bron}, E., {Pinto}, C., {et~al.} 2018, \aap, 615, A129

\bibitem[{{Jones} {et~al.}(2013){Jones}, {Fanciullo}, {K{\"o}hler},
  {Verstraete}, {Guillet}, {Bocchio}, \& {Ysard}}]{Jones2013}
{Jones}, A.~P., {Fanciullo}, L., {K{\"o}hler}, M., {et~al.} 2013, \aap, 558,
  A62

\bibitem[{{Jones} \& {Habart}(2015)}]{jones2015}
{Jones}, A.~P., \& {Habart}, E. 2015, \aap, 581, A92

\bibitem[{{Jones} {et~al.}(2017){Jones}, {K{\"o}hler}, {Ysard}, {Bocchio}, \&
  {Verstraete}}]{jones2017}
{Jones}, A.~P., {K{\"o}hler}, M., {Ysard}, N., {Bocchio}, M., \& {Verstraete},
  L. 2017, \aap, 602, A46

\bibitem[{{Juvela}(2019)}]{juvela2019}
{Juvela}, M. 2019, \aap, 622, A79

\bibitem[{{Kaplan} {et~al.}(2017){Kaplan}, {Dinerstein}, {Oh}, {Mace}, {Kim},
  {Sokal}, {Pavel}, {Lee}, {Pak}, {Park}, {Sok Oh}, \& {Jaffe}}]{Kaplan17}
{Kaplan}, K.~F., {Dinerstein}, H.~L., {Oh}, H., {et~al.} 2017, \apj, 838, 152

\bibitem[{{Kaufman} {et~al.}(2006){Kaufman}, {Wolfire}, \&
  {Hollenbach}}]{2006ApJ...644..283K}
{Kaufman}, M.~J., {Wolfire}, M.~G., \& {Hollenbach}, D.~J. 2006, \apj, 644, 283

\bibitem[{{Kaufman} {et~al.}(1999){Kaufman}, {Wolfire}, {Hollenbach}, \&
  {Luhman}}]{kaufman99}
{Kaufman}, M.~J., {Wolfire}, M.~G., {Hollenbach}, D.~J., \& {Luhman}, M.~L.
  1999, \apj, 527, 795

\bibitem[{{Kirsanova} \& {Wiebe}(2019)}]{Kirsanova19}
{Kirsanova}, M.~S., \& {Wiebe}, D.~S. 2019, \mnras, 486, 2525

\bibitem[{{Knight} {et~al.}(2021{\natexlab{a}}){Knight}, {Peeters}, {Stock},
  {Vacca}, \& {Tielens}}]{Knight21}
{Knight}, C., {Peeters}, E., {Stock}, D.~J., {Vacca}, W.~D., \& {Tielens},
  A.~G.~G.~M. 2021{\natexlab{a}}, \apj, 918, 8

\bibitem[{{Knight} {et~al.}(2021{\natexlab{b}}){Knight}, {Peeters}, {Tielens},
  \& {Vacca}}]{Knight21:Orion}
{Knight}, C., {Peeters}, E., {Tielens}, A.~G.~G.~M., \& {Vacca}, W.~D.
  2021{\natexlab{b}}, \mnras, doi:10.1093/mnras/stab3047

\bibitem[{{Kounkel} {et~al.}(2017){Kounkel}, {Hartmann}, {Loinard},
  {Ortiz-Le{\'o}n}, {Mioduszewski}, {Rodr{\'\i}guez}, {Dzib}, {Torres}, {Pech},
  {Galli}, {Rivera}, {Boden}, {Evans}, {Brice{\~n}o}, \& {Tobin}}]{Kounkel17}
{Kounkel}, M., {Hartmann}, L., {Loinard}, L., {et~al.} 2017, \apj, 834, 142

\bibitem[{{Lai} {et~al.}(2020){Lai}, {Smith}, {Baba}, {Spoon}, \&
  {Imanishi}}]{Lai:20}
{Lai}, T. S.~Y., {Smith}, J.~D.~T., {Baba}, S., {Spoon}, H. W.~W., \&
  {Imanishi}, M. 2020, \apj, 905, 55

\bibitem[{{Le Bourlot} {et~al.}(2012){Le Bourlot}, {Le Petit}, {Pinto},
  {Roueff}, \& {Roy}}]{lebourlot2012}
{Le Bourlot}, J., {Le Petit}, F., {Pinto}, C., {Roueff}, E., \& {Roy}, F. 2012,
  \aap, 541, A76

\bibitem[{{Le Bourlot} {et~al.}(1993){Le Bourlot}, {Pineau Des Forets},
  {Roueff}, \& {Flower}}]{lebourlot}
{Le Bourlot}, J., {Pineau Des Forets}, G., {Roueff}, E., \& {Flower}, D.~R.
  1993, A\&A, 267, 233

\bibitem[{{Le Petit} {et~al.}(2006){Le Petit}, {Nehm{\'e}}, {Le Bourlot}, \&
  {Roueff}}]{lepetit}
{Le Petit}, F., {Nehm{\'e}}, C., {Le Bourlot}, J., \& {Roueff}, E. 2006, ApJS,
  164, 506

\bibitem[{{Leger} \& {Puget}(1984)}]{leger_puget84}
{Leger}, A., \& {Puget}, J.~L. 1984, \aap, 500, 279

\bibitem[{{Leurini} {et~al.}(2006){Leurini}, {Rolffs}, {Thorwirth}, {Parise},
  {Schilke}, {Comito}, {Wyrowski}, {G{\"u}sten}, {Bergman}, {Menten}, \&
  {Nyman}}]{Leurini06}
{Leurini}, S., {Rolffs}, R., {Thorwirth}, S., {et~al.} 2006, \aap, 454, L47

\bibitem[{{Lis} \& {Schilke}(2003)}]{Lis03}
{Lis}, D.~C., \& {Schilke}, P. 2003, \apjl, 597, L145

\bibitem[{{Luhman} {et~al.}(1994){Luhman}, {Jaffe}, {Keller}, \&
  {Pak}}]{Luhman94}
{Luhman}, M.~L., {Jaffe}, D.~T., {Keller}, L.~D., \& {Pak}, S. 1994, \apjl,
  436, L185

\bibitem[{Malhotra {et~al.}(2001)Malhotra, Kaufman, Hollenbach, Helou, Rubin,
  Brauher, Dale, Lu, Lord, Stacey, {et~al.}}]{malhotra2001far}
Malhotra, S., Kaufman, M., Hollenbach, D., {et~al.} 2001, The Astrophysical
  Journal, 561, 766

\bibitem[{{Maltseva} {et~al.}(2015){Maltseva}, {Petrignani}, {Candian},
  {Mackie}, {Huang}, {Lee}, {Tielens}, {Oomens}, \& {Buma}}]{Maltseva:15}
{Maltseva}, E., {Petrignani}, A., {Candian}, A., {et~al.} 2015, \apj, 814, 23

\bibitem[{{Maragkoudakis} {et~al.}(2018){Maragkoudakis}, {Ivkovich}, {Peeters},
  {Stock}, {Hemachandra}, \& {Tielens}}]{Maragkoudakis:18}
{Maragkoudakis}, A., {Ivkovich}, N., {Peeters}, E., {et~al.} 2018, \mnras, 481,
  5370

\bibitem[{{Maragkoudakis} {et~al.}(2020){Maragkoudakis}, {Peeters}, \&
  {Ricca}}]{Maragkoudakis:20}
{Maragkoudakis}, A., {Peeters}, E., \& {Ricca}, A. 2020, \mnras, 494, 642

\bibitem[{{Marconi} {et~al.}(1998){Marconi}, {Testi}, {Natta}, \&
  {Walmsley}}]{Marconi98}
{Marconi}, A., {Testi}, L., {Natta}, A., \& {Walmsley}, C.~M. 1998, \aap, 330,
  696

\bibitem[{{Mart{\'\i}nez} {et~al.}(2020){Mart{\'\i}nez}, {Santoro}, {Merino},
  {Accolla}, {Lauwaet}, {Sobrado}, {Sabbah}, {Pelaez}, {Herrero}, {Tanarro},
  {Ag{\'u}ndez}, {Mart{\'\i}n-Jimenez}, {Otero}, {Ellis}, {Joblin},
  {Cernicharo}, \& {Mart{\'\i}n-Gago}}]{Martinez:20}
{Mart{\'\i}nez}, L., {Santoro}, G., {Merino}, P., {et~al.} 2020, Nature
  Astronomy, 4, 97

\bibitem[{{Mattioda} {et~al.}(2020){Mattioda}, {Hudgins}, {Boersma},
  {Bauschlicher}, {Ricca}, {Cami}, {Peeters}, {S{\'a}nchez de Armas}, {Puerta
  Saborido}, \& {Allamandola}}]{Mattioda:20}
{Mattioda}, A.~L., {Hudgins}, D.~M., {Boersma}, C., {et~al.} 2020, \apjs, 251,
  22

\bibitem[{McKinney {et~al.}(2020)McKinney, Pope, Armus, Chary,
  D{\'\i}az-Santos, Dickinson, \& Kirkpatrick}]{mck20}
McKinney, J., Pope, A., Armus, L., {et~al.} 2020, The Astrophysical Journal,
  892, 119

\bibitem[{{Meeus} {et~al.}(2013){Meeus}, {Salyk}, {Bruderer}, {Fedele},
  {Maaskant}, {Evans}, {van Dishoeck}, {Montesinos}, {Herczeg}, {Bouwman},
  {Green}, {Dominik}, {Henning}, \& {Vicente}}]{Meeus2013}
{Meeus}, G., {Salyk}, C., {Bruderer}, S., {et~al.} 2013, \aap, 559, A84

\bibitem[{{Menten} {et~al.}(2007){Menten}, {Reid}, {Forbrich}, \&
  {Brunthaler}}]{Menten07}
{Menten}, K.~M., {Reid}, M.~J., {Forbrich}, J., \& {Brunthaler}, A. 2007, \aap,
  474, 515

\bibitem[{Montillaud {et~al.}(2013)Montillaud, Joblin, \&
  Toublanc}]{montillaud2013}
Montillaud, J., Joblin, C., \& Toublanc, D. 2013, Astronomy \& Astrophysics,
  552, A15

\bibitem[{{Mori} {et~al.}(2014){Mori}, {Onaka}, {Sakon}, {Ishihara},
  {Shimonishi}, {Ohsawa}, \& {Bell}}]{Mori:14}
{Mori}, T.~I., {Onaka}, T., {Sakon}, I., {et~al.} 2014, \apj, 784, 53

\bibitem[{{Mori} {et~al.}(2012){Mori}, {Sakon}, {Onaka}, {Kaneda}, {Umehata},
  \& {Ohsawa}}]{Mori:12}
{Mori}, T.~I., {Sakon}, I., {Onaka}, T., {et~al.} 2012, \apj, 744, 68

\bibitem[{{Moutou} {et~al.}(1999){Moutou}, {Sellgren}, {Verstraete}, \& {L{\'
  e}ger}}]{moutou}
{Moutou}, C., {Sellgren}, K., {Verstraete}, L., \& {L{\' e}ger}, A. 1999, A\&A,
  347, 949

\bibitem[{{Murga} {et~al.}(2020){Murga}, {Kirsanova}, {Vasyunin}, \&
  {Pavlyuchenkov}}]{murga2020}
{Murga}, M.~S., {Kirsanova}, M.~S., {Vasyunin}, A.~I., \& {Pavlyuchenkov},
  Y.~N. 2020, \mnras, 497, 2327

\bibitem[{{Murga} {et~al.}(2022){Murga}, {Kirsanova}, {Wiebe}, \&
  {Boley}}]{murga2022}
{Murga}, M.~S., {Kirsanova}, M.~S., {Wiebe}, D.~S., \& {Boley}, P.~A. 2022,
  \mnras, 509, 800

\bibitem[{{Murga} {et~al.}(2019){Murga}, {Wiebe}, {Sivkova}, \&
  {Akimkin}}]{murga2019}
{Murga}, M.~S., {Wiebe}, D.~S., {Sivkova}, E.~E., \& {Akimkin}, V.~V. 2019,
  \mnras, 488, 965

\bibitem[{{Nagy} {et~al.}(2013){Nagy}, {Van der Tak}, {Ossenkopf}, {Gerin}, {Le
  Petit}, {Le Bourlot}, {Black}, {Goicoechea}, {Joblin}, {R{\"o}llig}, \&
  {Bergin}}]{Nagy13}
{Nagy}, Z., {Van der Tak}, F.~F.~S., {Ossenkopf}, V., {et~al.} 2013, \aap, 550,
  A96

\bibitem[{{Naslim} {et~al.}(2015){Naslim}, {Kemper}, {Madden}, {Hony}, {Chu},
  {Galliano}, {Bot}, {Yang}, {Seok}, {Oliveira}, {van Loon}, {Meixner}, {Li},
  {Hughes}, {Gordon}, {Otsuka}, {Hirashita}, {Morata}, {Lebouteiller},
  {Indebetouw}, {Srinivasan}, {Bernard}, \& {Reach}}]{Naslim15}
{Naslim}, N., {Kemper}, F., {Madden}, S.~C., {et~al.} 2015, \mnras, 446, 2490

\bibitem[{{Natta} {et~al.}(1994){Natta}, {Walmsley}, \& {Tielens}}]{Natta94}
{Natta}, A., {Walmsley}, C.~M., \& {Tielens}, A.~G.~G.~M. 1994, \apj, 428, 209

\bibitem[{{O'Dell}(2001)}]{Odell01}
{O'Dell}, C.~R. 2001, \araa, 39, 99

\bibitem[{{Onaka} {et~al.}(2014){Onaka}, {Mori}, {Sakon}, {Ohsawa}, {Kaneda},
  {Okada}, \& {Tanaka}}]{onaka2014}
{Onaka}, T., {Mori}, T.~I., {Sakon}, I., {et~al.} 2014, \apj, 780, 114

\bibitem[{{Ossenkopf} {et~al.}(2013){Ossenkopf}, {R{\"o}llig}, {Neufeld},
  {Pilleri}, {Lis}, {Fuente}, {van der Tak}, \& {Bergin}}]{Ossenkopf13}
{Ossenkopf}, V., {R{\"o}llig}, M., {Neufeld}, D.~A., {et~al.} 2013, \aap, 550,
  A57

\bibitem[{Osterbrock \& Ferland(2006)}]{osterbrock2006astrophysics}
Osterbrock, D.~E., \& Ferland, G.~J. 2006, Astrophysics Of Gas Nebulae and
  Active Galactic Nuclei (University science books)

\bibitem[{{Pabst} {et~al.}(2019){Pabst}, {Higgins}, {Goicoechea}, {Teyssier},
  {Bern{\'e}}, {Chambers}, {Wolfire}, {Suri}, {Guesten}, {Stutzki}, {Graf},
  {Risacher}, \& {Tielens}}]{Pabst19}
{Pabst}, C., {Higgins}, R., {Goicoechea}, J.~R., {et~al.} 2019, \nat, 565, 618

\bibitem[{{Pabst} {et~al.}(2017){Pabst}, {Goicoechea}, {Teyssier}, {Bern{\'e}},
  {Ochsendorf}, {Wolfire}, {Higgins}, {Riquelme}, {Risacher}, {Pety}, {Le
  Petit}, {Roueff}, {Bron}, \& {Tielens}}]{Pabst17}
{Pabst}, C.~H.~M., {Goicoechea}, J.~R., {Teyssier}, D., {et~al.} 2017, \aap,
  606, A29

\bibitem[{{Pabst} {et~al.}(2020){Pabst}, {Goicoechea}, {Teyssier}, {Bern{\'e}},
  {Higgins}, {Chambers}, {Kabanovic}, {G{\"u}sten}, {Stutzki}, \&
  {Tielens}}]{Pabst20}
---. 2020, \aap, 639, A2

\bibitem[{{Paladini} {et~al.}(2012){Paladini}, {Umana}, {Veneziani},
  {Noriega-Crespo}, {Anderson}, {Piacentini}, {Pinheiro Gon{\c{c}}alves},
  {Paradis}, {Tibbs}, {Bernard}, \& {Natoli}}]{Paladini:12}
{Paladini}, R., {Umana}, G., {Veneziani}, M., {et~al.} 2012, \apj, 760, 149

\bibitem[{Parker {et~al.}(2012)Parker, Zhang, Kim, Kaiser, Landera, Kislov,
  Mebel, \& Tielens}]{Parker2012}
Parker, D. S.~N., Zhang, F., Kim, Y.~S., {et~al.} 2012, Proceedings of the
  National Academy of Sciences, 109, 53.
\newblock \url{https://www.pnas.org/content/109/1/53}

\bibitem[{{Parmar} {et~al.}(1991){Parmar}, {Lacy}, \& {Achtermann}}]{Parmar91}
{Parmar}, P.~S., {Lacy}, J.~H., \& {Achtermann}, J.~M. 1991, \apjl, 372, L25

\bibitem[{{Parravano} {et~al.}(2003){Parravano}, {Hollenbach}, \&
  {McKee}}]{Parravano03}
{Parravano}, A., {Hollenbach}, D.~J., \& {McKee}, C.~F. 2003, \apj, 584, 797

\bibitem[{{Peeters} {et~al.}(2004{\natexlab{a}}){Peeters}, {Allamandola},
  {Bauschlicher}, {Hudgins}, {Sandford}, \& {Tielens}}]{Peeters04}
{Peeters}, E., {Allamandola}, L.~J., {Bauschlicher}, C.~W., J., {et~al.}
  2004{\natexlab{a}}, \apj, 604, 252

\bibitem[{{Peeters} {et~al.}(2017){Peeters}, {Bauschlicher}, {Allamandola},
  {Tielens}, {Ricca}, \& {Wolfire}}]{peeters17}
{Peeters}, E., {Bauschlicher}, Jr., C.~W., {Allamandola}, L.~J., {et~al.} 2017,
  ApJ, 836, 198

\bibitem[{{Peeters} {et~al.}(2002){Peeters}, {Hony}, {Van Kerckhoven},
  {Tielens}, {Allamandola}, {Hudgins}, \& {Bauschlicher}}]{Peeters:prof6:02}
{Peeters}, E., {Hony}, S., {Van Kerckhoven}, C., {et~al.} 2002, \aap, 390, 1089

\bibitem[{{Peeters} {et~al.}(2004{\natexlab{b}}){Peeters}, {Spoon}, \&
  {Tielens}}]{Peeters:pahtracer:04}
{Peeters}, E., {Spoon}, H.~W.~W., \& {Tielens}, A.~G.~G.~M. 2004{\natexlab{b}},
  \apj, 613, 986

\bibitem[{{Peeters} {et~al.}(2012){Peeters}, {Tielens}, {Allamandola}, \&
  {Wolfire}}]{Peeters:12}
{Peeters}, E., {Tielens}, A.~G.~G.~M., {Allamandola}, L.~J., \& {Wolfire},
  M.~G. 2012, \apj, 747, 44

\bibitem[{{Pellegrini} {et~al.}(2009){Pellegrini}, {Baldwin}, {Ferland},
  {Shaw}, \& {Heathcote}}]{Pellegrini09}
{Pellegrini}, E.~W., {Baldwin}, J.~A., {Ferland}, G.~J., {Shaw}, G., \&
  {Heathcote}, S. 2009, \apj, 693, 285

\bibitem[{{Pilleri} {et~al.}(2015){Pilleri}, {Joblin}, {Boulanger}, \&
  {Onaka}}]{pilleri2015}
{Pilleri}, P., {Joblin}, C., {Boulanger}, F., \& {Onaka}, T. 2015, \aap, 577,
  A16

\bibitem[{{Pilleri} {et~al.}(2012){Pilleri}, {Montillaud}, {Bern{\'e}}, \&
  {Joblin}}]{pilleri12}
{Pilleri}, P., {Montillaud}, J., {Bern{\'e}}, O., \& {Joblin}, C. 2012, A\&A,
  542, A69

\bibitem[{{Pilleri} {et~al.}(2013){Pilleri}, {Trevi{\~n}o-Morales}, {Fuente},
  {Joblin}, {Cernicharo}, {Gerin}, {Viti}, {Bern{\'e}}, {Goicoechea}, {Pety},
  {Gonzalez-Garc{\'\i}a}, {Montillaud}, {Ossenkopf}, {Kramer},
  {Garc{\'\i}a-Burillo}, {Le Petit}, \& {Le Bourlot}}]{Pilleri2013}
{Pilleri}, P., {Trevi{\~n}o-Morales}, S., {Fuente}, A., {et~al.} 2013, \aap,
  554, A87

\bibitem[{{Pound} \& {Wolfire}(2008)}]{2008ASPC..394..654P}
{Pound}, M.~W., \& {Wolfire}, M.~G. 2008, in Astronomical Society of the
  Pacific Conference Series, Vol. 394, Astronomical Data Analysis Software and
  Systems XVII, ed. R.~W. {Argyle}, P.~S. {Bunclark}, \& J.~R. {Lewis}, 654

\bibitem[{{Povich} {et~al.}(2007){Povich}, {Stone}, {Churchwell}, {Zweibel},
  {Wolfire}, {Babler}, {Indebetouw}, {Meade}, \& {Whitney}}]{Povich:07}
{Povich}, M.~S., {Stone}, J.~M., {Churchwell}, E., {et~al.} 2007, \apj, 660,
  346

\bibitem[{{Putaud} {et~al.}(2019){Putaud}, {Michaut}, {Le Petit}, {Roueff}, \&
  {Lis}}]{Putaud19}
{Putaud}, T., {Michaut}, X., {Le Petit}, F., {Roueff}, E., \& {Lis}, D.~C.
  2019, \aap, 632, A8

\bibitem[{{Rapacioli} {et~al.}(2006){Rapacioli}, {Calvo}, {Joblin}, {Parneix},
  {Toublanc}, \& {Spiegelman}}]{Rapacioli2006}
{Rapacioli}, M., {Calvo}, F., {Joblin}, C., {et~al.} 2006, \aap, 460, 519

\bibitem[{Rapacioli {et~al.}(2005)Rapacioli, Joblin, \&
  Boissel}]{rapacioli2005}
Rapacioli, M., Joblin, C., \& Boissel, P. 2005, Astronomy \& Astrophysics, 429,
  193

\bibitem[{{Reach} {et~al.}(2005){Reach}, {Megeath}, {Cohen}, {Hora}, {Carey},
  {Surace}, {Willner}, {Barmby}, {Wilson}, {Glaccum}, {Lowrance}, {Marengo}, \&
  {Fazio}}]{reach}
{Reach}, W.~T., {Megeath}, S.~T., {Cohen}, M., {et~al.} 2005, PASP, 117, 978

\bibitem[{{Ricca} {et~al.}(2012){Ricca}, {Bauschlicher}, {Boersma}, {Tielens},
  \& {Allamandola}}]{Ricca12}
{Ricca}, A., {Bauschlicher}, Charles~W., J., {Boersma}, C., {Tielens}, A.
  G.~G.~M., \& {Allamandola}, L.~J. 2012, \apj, 754, 75

\bibitem[{Rieke {et~al.}(2015)Rieke, Wright, B{\"o}ker, Bouwman, Colina,
  Glasse, Gordon, Greene, G{\"u}del, Henning, {et~al.}}]{rieke15}
Rieke, G.~H., Wright, G., B{\"o}ker, T., {et~al.} 2015, Publications of the
  Astronomical Society of the Pacific, 127, 584

\bibitem[{{Robberto} {et~al.}(2020){Robberto}, {Gennaro}, {Ubeira Gabellini},
  {Hillenbrand}, {Pacifici}, {Ubeda}, {Andersen}, {Barman}, {Bellini}, {Da
  Rio}, {de Mink}, {Lodato}, {Manara}, {Platais}, {Pueyo}, {Strampelli}, {Tan},
  \& {Testi}}]{rob20}
{Robberto}, M., {Gennaro}, M., {Ubeira Gabellini}, M.~G., {et~al.} 2020, \apj,
  896, 79

\bibitem[{{R{\"o}llig} {et~al.}(2013){R{\"o}llig}, {Szczerba}, {Ossenkopf}, \&
  {Gl{\"u}ck}}]{markus}
{R{\"o}llig}, M., {Szczerba}, R., {Ossenkopf}, V., \& {Gl{\"u}ck}, C. 2013,
  A\&A, 549, A85

\bibitem[{{R{\"o}llig} {et~al.}(2007){R{\"o}llig}, {Abel}, {Bell}, {Bensch},
  {Black}, {Ferland}, {Jonkheid}, { }, {Kaufman}, {Le Bourlot}, {Le Petit},
  {Meijerink}, {Morata}, {Ossenkopf}, {Roueff}, {Shaw}, {Spaans}, {Sternberg},
  {Stutzki}, {Thi}, {van Dishoeck}, {van Hoof}, {Viti}, \&
  {Wolfire}}]{comparison07}
{R{\"o}llig}, M., {Abel}, N.~P., {Bell}, T., {et~al.} 2007, \aap, 467, 187

\bibitem[{{Rosenthal} {et~al.}(2000){Rosenthal}, {Bertoldi}, \&
  {Drapatz}}]{Rosenthal2000}
{Rosenthal}, D., {Bertoldi}, F., \& {Drapatz}, S. 2000, \aap, 356, 705

\bibitem[{{Sabbah} {et~al.}(2017){Sabbah}, {Bonnamy}, {Papanastasiou},
  {Cernicharo}, {Mart{\'\i}n-Gago}, \& {Joblin}}]{Sabbah:17}
{Sabbah}, H., {Bonnamy}, A., {Papanastasiou}, D., {et~al.} 2017, \apj, 843, 34

\bibitem[{{Salama} {et~al.}(2018){Salama}, {Sciamma-O'Brien}, {Contreras}, \&
  {Bejaoui}}]{Salama:18}
{Salama}, F., {Sciamma-O'Brien}, E., {Contreras}, C.~S., \& {Bejaoui}, S. 2018,
  IAU Symposium, 332, 364

\bibitem[{{Salgado} {et~al.}(2016){Salgado}, {Bern{\'e}}, {Adams}, {Herter},
  {Keller}, \& {Tielens}}]{Salgado16}
{Salgado}, F., {Bern{\'e}}, O., {Adams}, J.~D., {et~al.} 2016, \apj, 830, 118

\bibitem[{Schirmer {et~al.}(2020)Schirmer, Abergel, Verstraete, Ysard, Juvela,
  Jones, \& Habart}]{schirmer2020}
Schirmer, T., Abergel, A., Verstraete, L., {et~al.} 2020, Astronomy \&
  Astrophysics, 639, A144

\bibitem[{Schirmer {et~al.}(in prep)Schirmer, Ysard, Jones, Abergel, Habart, \&
  Verstraete}]{Schirmer2022}
Schirmer, T., Ysard, N., Jones, A., {et~al.} in prep

\bibitem[{{Schirmer} {et~al.}(2021){Schirmer}, {Habart}, {Ysard}, {Bron}, {Le
  Bourlot}, {Verstraete}, {Abergel}, {Jones}, {Roueff}, \& {Le
  Petit}}]{schirmer2021}
{Schirmer}, T., {Habart}, E., {Ysard}, N., {et~al.} 2021, \aap, 649, A148

\bibitem[{{Sciamma-O'Brien} \& {Salama}(2020)}]{Sciamma:20}
{Sciamma-O'Brien}, E., \& {Salama}, F. 2020, \apj, 905, 45

\bibitem[{{Sellgren} {et~al.}(1990){Sellgren}, {Tokunaga}, \&
  {Nakada}}]{Sellgren90}
{Sellgren}, K., {Tokunaga}, A.~T., \& {Nakada}, Y. 1990, \apj, 349, 120

\bibitem[{Sellgren {et~al.}(2010)Sellgren, Werner, Ingalls, Smith, Carleton, \&
  Joblin}]{sellgren2010c60}
Sellgren, K., Werner, M.~W., Ingalls, J.~G., {et~al.} 2010, The Astrophysical
  Journal Letters, 722, L54

\bibitem[{{Shannon} \& {Boersma}(2018)}]{Shannon:18}
{Shannon}, M.~J., \& {Boersma}, C. 2018, in {P}roceedings of the 17th {P}ython
  in {S}cience {C}onference, ed. F.~{Akici}, D.~{Lippa}, D.~{Niederhut}, \&
  M.~{Pacer}, 99

\bibitem[{{Shaw} {et~al.}(2009){Shaw}, {Ferland}, {Henney}, {Stancil}, {Abel},
  {Pellegrini}, {Baldwin}, \& {van Hoof}}]{Shaw09}
{Shaw}, G., {Ferland}, G.~J., {Henney}, W.~J., {et~al.} 2009, \apj, 701, 677

\bibitem[{Sheffer {et~al.}(2011)Sheffer, Wolfire, Hollenbach, Kaufman, \&
  Cordier}]{Sheffer11}
Sheffer, Y., Wolfire, M.~G., Hollenbach, D.~J., Kaufman, M.~J., \& Cordier, M.
  2011, ApJ, 741, 45

\bibitem[{{Sidhu} {et~al.}(2021){Sidhu}, {Peeters}, {Cami}, \&
  {Knight}}]{Sidhu:21}
{Sidhu}, A., {Peeters}, E., {Cami}, J., \& {Knight}, C. 2021, \mnras, 500, 177

\bibitem[{{Simon} {et~al.}(1997){Simon}, {Stutzki}, {Sternberg}, \&
  {Winnewisser}}]{Simon97}
{Simon}, R., {Stutzki}, J., {Sternberg}, A., \& {Winnewisser}, G. 1997, \aap,
  327, L9

\bibitem[{{Sloan} {et~al.}(1997){Sloan}, {Bregman}, {Geballe}, {Allamandola},
  \& {Woodward}}]{Sloan:97}
{Sloan}, G.~C., {Bregman}, J.~D., {Geballe}, T.~R., {Allamandola}, L.~J., \&
  {Woodward}, E. 1997, \apj, 474, 735

\bibitem[{{Smith} {et~al.}(2007){Smith}, {Armus}, {Dale}, {Roussel}, {Sheth},
  {Buckalew}, {Jarrett}, {Helou}, \& {Kennicutt}}]{smith}
{Smith}, J.~D.~T., {Armus}, L., {Dale}, D.~A., {et~al.} 2007, \pasp, 119, 1133

\bibitem[{{Stepnik} {et~al.}(2003){Stepnik}, {Abergel}, {Bernard}, {Boulanger},
  {Cambr{\'e}sy}, {Giard}, {Jones}, {Lagache}, {Lamarre}, {Meny}, {Pajot}, {Le
  Peintre}, {Ristorcelli}, {Serra}, \& {Torre}}]{stepnik2003}
{Stepnik}, B., {Abergel}, A., {Bernard}, J.~P., {et~al.} 2003, \aap, 398, 551

\bibitem[{{Sternberg} \& {Dalgarno}(1989{\natexlab{a}})}]{stenberg}
{Sternberg}, A., \& {Dalgarno}, A. 1989{\natexlab{a}}, \apj, 338, 197

\bibitem[{{Sternberg} \& {Dalgarno}(1989{\natexlab{b}})}]{sternberg}
---. 1989{\natexlab{b}}, ApJ, 338, 197

\bibitem[{{Sternberg} \& {Dalgarno}(1995)}]{Sternberg95}
---. 1995, \apjs, 99, 565

\bibitem[{{Stock} {et~al.}(2016){Stock}, {Choi}, {Moya}, {Otaguro}, {Sorkhou},
  {Allamandola}, {Tielens}, \& {Peeters}}]{Stock:16}
{Stock}, D.~J., {Choi}, W.~D.~Y., {Moya}, L.~G.~V., {et~al.} 2016, \apj, 819,
  65

\bibitem[{{Stock} \& {Peeters}(2017)}]{Stock2017}
{Stock}, D.~J., \& {Peeters}, E. 2017, \apj, 837, 129

\bibitem[{{Stoerzer} {et~al.}(1995){Stoerzer}, {Stutzki}, \&
  {Sternberg}}]{Stoerzer95}
{Stoerzer}, H., {Stutzki}, J., \& {Sternberg}, A. 1995, \aap, 296, L9

\bibitem[{{St{\"o}rzer} \& {Hollenbach}(1998)}]{Stoerzer98}
{St{\"o}rzer}, H., \& {Hollenbach}, D. 1998, \apj, 495, 853

\bibitem[{{Tabone} {et~al.}(2021){Tabone}, {van Hemert}, {van Dishoeck}, \&
  {Black}}]{Tabone2021}
{Tabone}, B., {van Hemert}, M.~C., {van Dishoeck}, E.~F., \& {Black}, J.~H.
  2021, \aap, 650, A192

\bibitem[{{Tauber} {et~al.}(1995){Tauber}, {Lis}, {Keene}, {Schilke}, \&
  {Buettgenbach}}]{Tauber95}
{Tauber}, J.~A., {Lis}, D.~C., {Keene}, J., {Schilke}, P., \& {Buettgenbach},
  T.~H. 1995, \aap, 297, 567

\bibitem[{{Tauber} {et~al.}(1994){Tauber}, {Tielens}, {Meixner}, \&
  {Goldsmith}}]{Tauber94}
{Tauber}, J.~A., {Tielens}, A.~G.~G.~M., {Meixner}, M., \& {Goldsmith}, P.~F.
  1994, \apj, 422, 136

\bibitem[{Tielens(2005)}]{Tielens_book05}
Tielens, A. 2005, The Physics and Chemistry of the Interstellar Medium

\bibitem[{{Tielens} \& {Hollenbach}(1985{\natexlab{a}})}]{tielens:85b}
{Tielens}, A.~G.~G.~M., \& {Hollenbach}, D. 1985{\natexlab{a}}, ApJ, 291, 747

\bibitem[{{Tielens} \& {Hollenbach}(1985{\natexlab{b}})}]{tielens:85}
---. 1985{\natexlab{b}}, ApJ, 291, 722

\bibitem[{{Tielens} {et~al.}(1993){Tielens}, {Meixner}, {van der Werf},
  {Bregman}, {Tauber}, {Stutzki}, \& {Rank}}]{Tielens93}
{Tielens}, A.~G.~G.~M., {Meixner}, M.~M., {van der Werf}, P.~P., {et~al.} 1993,
  Science, 262, 86

\bibitem[{{van den Ancker} {et~al.}(2000){van den Ancker}, {Tielens}, \&
  {Wesselius}}]{vandenAncker:00}
{van den Ancker}, M.~E., {Tielens}, A.~G.~G.~M., \& {Wesselius}, P.~R. 2000,
  \aap, 358, 1035

\bibitem[{{van der Tak} {et~al.}(2013){van der Tak}, {Nagy}, {Ossenkopf},
  {Makai}, {Black}, {Faure}, {Gerin}, \& {Bergin}}]{Tak13}
{van der Tak}, F.~F.~S., {Nagy}, Z., {Ossenkopf}, V., {et~al.} 2013, \aap, 560,
  A95

\bibitem[{{van der Werf} {et~al.}(2013){van der Werf}, {Goss}, \&
  {O'Dell}}]{vanderWerf13}
{van der Werf}, P.~P., {Goss}, W.~M., \& {O'Dell}, C.~R. 2013, \apj, 762, 101

\bibitem[{{van der Werf} {et~al.}(1996){van der Werf}, {Stutzki}, {Sternberg},
  \& {Krabbe}}]{vanderWerf96}
{van der Werf}, P.~P., {Stutzki}, J., {Sternberg}, A., \& {Krabbe}, A. 1996,
  \aap, 313, 633

\bibitem[{{van Dishoeck} \& {Black}(1986)}]{vandishoeck86}
{van Dishoeck}, E.~F., \& {Black}, J.~H. 1986, \apjs, 62, 109

\bibitem[{Verma {et~al.}(2003)Verma, Lutz, Sturm, Sternberg, Genzel, \&
  Vacca}]{verma2003mid}
Verma, A., Lutz, D., Sturm, E., {et~al.} 2003, Astronomy \& Astrophysics, 403,
  829

\bibitem[{{Verstraete} {et~al.}(1990){Verstraete}, {Leger}, {D'Hendecourt},
  {Defourneau}, \& {Dutuit}}]{Verstraete1990}
{Verstraete}, L., {Leger}, A., {D'Hendecourt}, L., {Defourneau}, D., \&
  {Dutuit}, O. 1990, \aap, 237, 436

\bibitem[{Visser {et~al.}(2007)Visser, Geers, Dullemond, Augereau, Pontoppidan,
  \& van Dishoeck}]{vis07}
Visser, R., Geers, V., Dullemond, C., {et~al.} 2007, Astronomy \& Astrophysics,
  466, 229

\bibitem[{{Wakelam} {et~al.}(2017){Wakelam}, {Bron}, {Cazaux}, {Dulieu}, {Gry},
  {Guillard}, {Habart}, {Hornek{\ae}r}, {Morisset}, {Nyman}, {Pirronello},
  {Price}, {Valdivia}, {Vidali}, \& {Watanabe}}]{Wakelam2017}
{Wakelam}, V., {Bron}, E., {Cazaux}, S., {et~al.} 2017, Molecular Astrophysics,
  9, 1

\bibitem[{Walmsley {et~al.}(2000)Walmsley, Natta, Oliva, \&
  Testi}]{walmsley2000structure}
Walmsley, C., Natta, A., Oliva, E., \& Testi, L. 2000, Astronomy and
  Astrophysics, 364, 301

\bibitem[{{Walmsley} {et~al.}(2000){Walmsley}, {Natta}, {Oliva}, \&
  {Testi}}]{Walmsley00}
{Walmsley}, C.~M., {Natta}, A., {Oliva}, E., \& {Testi}, L. 2000, \aap, 364,
  301

\bibitem[{{Watson} {et~al.}(2008){Watson}, {Povich}, {Churchwell}, {Babler},
  {Chunev}, {Hoare}, {Indebetouw}, {Meade}, {Robitaille}, \&
  {Whitney}}]{Watson:08}
{Watson}, C., {Povich}, M.~S., {Churchwell}, E.~B., {et~al.} 2008, \apj, 681,
  1341

\bibitem[{{Weilbacher} {et~al.}(2015){Weilbacher}, {Monreal-Ibero},
  {Kollatschny}, {Ginsburg}, {McLeod}, {Kamann}, {Sandin}, {Palsa}, {Wisotzki},
  {Bacon}, {Selman}, {Brinchmann}, {Caruana}, {Kelz}, {Martinsson},
  {P{\'e}contal-Rousset}, {Richard}, \& {Wendt}}]{Weilbacher15}
{Weilbacher}, P.~M., {Monreal-Ibero}, A., {Kollatschny}, W., {et~al.} 2015,
  \aap, 582, A114

\bibitem[{Weingartner \& Draine(2001)}]{weingartner_photoelectric_2001}
Weingartner, J.~C., \& Draine, B.~T. 2001, The Astrophysical Journal Supplement
  Series, 134, 263.
\newblock \url{http://stacks.iop.org/0067-0049/134/i=2/a=263}

\bibitem[{Wen \& O'dell(1995)}]{wen1995three}
Wen, Z., \& O'dell, C. 1995, The Astrophysical Journal, 438, 784

\bibitem[{{Werner} {et~al.}(1976){Werner}, {Gatley}, {Harper}, {Becklin},
  {Loewenstein}, {Telesco}, \& {Thronson}}]{Werner76}
{Werner}, M.~W., {Gatley}, I., {Harper}, D.~A., {et~al.} 1976, \apj, 204, 420

\bibitem[{{Wiersma} {et~al.}(2020){Wiersma}, {Candian}, {Bakker}, {Martens},
  {Berden}, {Oomens}, {Buma}, \& {Petrignani}}]{Wiersma:20}
{Wiersma}, S.~D., {Candian}, A., {Bakker}, J.~M., {et~al.} 2020, \aap, 635, A9

\bibitem[{Woitke {et~al.}(2009)Woitke, Kamp, \& Thi}]{woitke2009radiation}
Woitke, P., Kamp, I., \& Thi, W.-F. 2009, Astronomy \& Astrophysics, 501, 383

\bibitem[{Wolfire {et~al.}(2003)Wolfire, McKee, Hollenbach, \&
  Tielens}]{wolfire_neutral_2003}
Wolfire, M.~G., McKee, C.~F., Hollenbach, D., \& Tielens, A. G. G.~M. 2003, The
  Astrophysical Journal, 587, 278.
\newblock \url{http://stacks.iop.org/0004-637X/587/i=1/a=278}

\bibitem[{{Wolfire} {et~al.}(1990){Wolfire}, {Tielens}, \&
  {Hollenbach}}]{wolfire90}
{Wolfire}, M.~G., {Tielens}, A.~G.~G.~M., \& {Hollenbach}, D. 1990, \apj, 358,
  116

\bibitem[{{Wright} {et~al.}(1999){Wright}, {van Dishoeck}, {Cox}, {Sidher}, \&
  {Kessler}}]{Wright99}
{Wright}, C.~M., {van Dishoeck}, E.~F., {Cox}, P., {Sidher}, S.~D., \&
  {Kessler}, M.~F. 1999, \apjl, 515, L29

\bibitem[{{Wyrowski} {et~al.}(1997){Wyrowski}, {Schilke}, {Hofner}, \&
  {Walmsley}}]{Wyrowski97}
{Wyrowski}, F., {Schilke}, P., {Hofner}, P., \& {Walmsley}, C.~M. 1997, \apjl,
  487, L171

\bibitem[{{Young Owl} {et~al.}(2000){Young Owl}, {Meixner}, {Wolfire},
  {Tielens}, \& {Tauber}}]{YoungOwl00}
{Young Owl}, R.~C., {Meixner}, M.~M., {Wolfire}, M., {Tielens}, A.~G.~G.~M., \&
  {Tauber}, J. 2000, \apj, 540, 886

\bibitem[{Zettergren {et~al.}({2021})Zettergren, Domaracka, Schlathoelter,
  Bolognesi, Diaz-Tendero, Labuda, Tosic, Maclot, Johnsson, Steber, Tikhonov,
  Castrovilli, Avaldi, Bari, Milosavljevic, Palacios, Faraji, Piekarski,
  Rousseau, Ascenzi, Romanzin, Erdmann, Alcami, Kopyra, Limao-Vieira, Kocisek,
  Fedor, Albertini, Gatchell, Cederquist, Schmidt, Gruber, Andersen, Heber,
  Toker, Hansen, Noble, Jouvet, Kjaer, Nielsen, Carrascosa, Bull, Candian, \&
  Petrignani}]{Zettergen:21}
Zettergren, H., Domaracka, A., Schlathoelter, T., {et~al.} {2021}, {EUROPEAN
  PHYSICAL JOURNAL D}, {75}, doi:{10.1140/epjd/s10053-021-00155-y}

\bibitem[{{Zhang} {et~al.}(2021){Zhang}, {Cummings}, {Wan}, {Yang}, \&
  {Stancil}}]{Zhang21}
{Zhang}, Z.~E., {Cummings}, S.~J., {Wan}, Y., {Yang}, B., \& {Stancil}, P.~C.
  2021, \apj, 912, 116

\end{thebibliography}

\end{document}